\documentclass[a4paper,11pt]{article}
\usepackage{jheppub} 
\usepackage{lineno}
\usepackage{float}
\usepackage{booktabs}
\usepackage{cleveref}
\usepackage[table]{xcolor}

\usepackage{tikz,hyperref}

\definecolor{lime}{HTML}{A6CE39}
\DeclareRobustCommand{\orcidicon}{\hspace{-1mm}
	\begin{tikzpicture}
		\draw[lime, fill=lime] (0,0) 
		circle [radius=0.12] 
		node[white] {{\fontfamily{qag}\selectfont \tiny \,ID}};
		\draw[white, fill=white] (-0.0525,0.095) 
		circle [radius=0.007];
	\end{tikzpicture}
	\hspace{-3mm}
}

\foreach \x in {A, ..., Z}{\expandafter\xdef\csname orcid\x\endcsname{\noexpand\href{https://orcid.org/\csname orcidauthor\x\endcsname}
		{\noexpand\orcidicon}}
}

\title{Exploring Quantumness at Long-Baseline Neutrino Experiments}

\author[a]{Murshed Alam\orcidA,}
\author[a]{Vedran Brdar\orcidB,}
\author[a]{and Dibya S. Chattopadhyay\orcidC}

\affiliation[a]{Department of Physics, Oklahoma State University, Stillwater, OK 74078, USA}

\emailAdd{murshed.alam@okstate.edu}
\emailAdd{vedran.brdar@okstate.edu}
\emailAdd{dibya.chattopadhyay@okstate.edu}

\abstract{
Violations of classicality can be probed through measurements performed on a system at different times, as proposed by Leggett and Garg.
Specifically, violations of Leggett-Garg inequalities suggest the presence of quantum effects in macroscopic systems.
Long-baseline neutrino experiments provide some of the longest available propagation distances over which such tests can be performed.
Previous studies of Leggett-Garg tests in the neutrino sector have largely focused on showing that the oscillation probabilities can violate classical bounds for certain parameter choices.
In this work, we develop a more complete and data-driven framework that treats both the distributions representing the classical and quantum behavior, as well as the experimental uncertainties. 
We consider MINOS, T2K, NOvA, as well as the upcoming DUNE, and present the respective statistical significance for distinguishing quantum behavior from classical scenarios at these long-baseline neutrino experiments.
Among them, we find that T2K yields the most significant violation of classicality, at the level of $\sim \! 14\sigma$, with NOvA and projections for DUNE also resulting in a significance of more than $5\sigma$.}

\begin{document}
\maketitle
\flushbottom

\section{Introduction}
\label{sec:intro}

Quantum mechanics has been established for about a century and today stands at the forefront of modern research. Its principles govern phenomena describing structure of matter and the behavior of elementary particles, which is far beyond the realm of classical mechanics. It is, however, nontrivial to set up a test to determine whether a system behaves in a classical or a quantum way. Among the most studied examples for that are Bell's inequalities~\cite{Bell:1964kc}, which quantify the correlations between measurements on spatially separated systems and provide a way to test whether these correlations require quantum mechanics for their explanation~\cite{Freedman:1972zza, Aspect:1982fx}.

Another probe of quantumness, which can be applied to a single system that is measured at different times, was formulated by Leggett and Garg~\cite{Leggett:1985zz, Emary:2013wfl}.
It attempts to test two fundamental classical criteria: macroscopic realism \cite{Leggett:1985zz, Leggett:2002ifn, Kofler:2007haj, Emary:2013wfl} and non-invasive measurability \cite{Emary:2013wfl, Fritz:2010qzm, Kofler:2008ubd, PhysRevLett.64.2358, PhysRevLett.63.2159, Wilde:2011aev, Fritz:2010qzm}. 
The former suggests that measurements reveal pre-existing values, while the latter asserts that measurements can be made without altering the state of the system. 
Correlations between measurements on a system at different times are bounded in a purely classical scenario, which is quantified by the Leggett–Garg inequality (LGI). Violations of this inequality therefore suggest that the system in question possesses quantum mechanical properties.

Long-baseline neutrino experiments provide some of the longest length scales over which tests of quantumness can be performed.
It was pioneered in Ref.~\cite{Formaggio:2016cuh} that the LGI can be tested in such experiments.
In that work, an additional assumption, called stationarity, was made: correlations between measurements are assumed to depend on the durations of the intervals rather than on the absolute times at which the measurements are performed. 
The authors employed data from the MINOS experiment \cite{Sousa:2015bxa, Holin:2015lya} and found that the LGI is violated at more than $6\sigma$.
This claim was recently challenged in Ref.~\cite{Groth:2025gtf}, where the authors used modified classical predictions and concluded that LGI violations at MINOS can be quantified only at the $\sim \! 3\sigma$ level. Since the jury is still out on this, in the present work, we obtain results using approaches from both Refs.~\cite{Formaggio:2016cuh, Groth:2025gtf}.
Our main improvement over both, however, lies in the treatment of the width of the experimental energy bins, which are important for creating data sequences used in testing LGI violations.
We also introduce a sampling method that handles asymmetric uncertainties on the neutrino survival probability data, and an effective cumulative distribution function (CDF)-based Gaussian fit that is employed for scenarios where a non-Gaussian shape of the classical predictions arise.
On the theory side, we treat the two classical prediction schemes~\cite{Formaggio:2016cuh, Groth:2025gtf} and introduce a modified RMS (root-mean-squared) $z$-score measure for quantifying the degree of non-classicality; our modified RMS $z$-score definition allows us to probe both the effects of the fraction of data points that violates the LGI, as well as the magnitude of the LGI violation.

In addition to MINOS, we also, for the first time, present results for two leading long-baseline experiments, T2K \cite{T2K:2011qtm} and NOvA\footnote{Note that Ref.~\cite{Barrios:2023yub} previously performed LGI tests for the NOvA experiment by evaluating the LG measures $K_3$ and $K_4$ to provide a qualitative demonstration of the violation of classical realism for NOvA. In this paper, we go a step further by quantifying the significance of the quantum behavior for NOvA.}~\cite{NOvA:2019cyt}, as well as projections for the upcoming DUNE experiment \cite{DUNE:2020ypp, DUNE:2020jqi}.
While it has been shown for several neutrino experiments that one would expect a violation of the classical limit in a certain parameter range \cite{Gangopadhyay:2013aha, Gangopadhyay:2017nsn, Naikoo:2019eec, Bittencourt:2022tcl}, detailed statistical analysis yielding the significance of LGI violation has only been presented in a few works, including the above Refs.~\cite{Formaggio:2016cuh, Groth:2025gtf} for MINOS and Refs.~\cite{Fu:2017hky, Wang:2022tnr}, which featured studies of the Daya Bay \cite{DayaBay:2012fng} and KamLAND \cite{KamLAND:2002uet} reactor antineutrino experiments. In this work, we address this gap by performing a detailed study of present and near-future accelerator-based long-baseline neutrino experiments in the context of LGI violation and ranking these experiments according to their potential for testing it.

The paper is organized as follows. In \cref{sec:Formalism}, we introduce the formalism behind Leggett-Garg (LG) strings and discuss how LGI violations can be tested in neutrino oscillation experiments, focusing on a treatment of both the quantum and classical predictions.
Further, we also introduce a modified RMS $z$-score definition and discuss why it is a well-suited measure for quantifying quantumness at neutrino experiments.
In \cref{sec:method}, we detail our analysis, with discussions on the effects of the width of the experimental energy bins and asymmetric uncertainties often present in neutrino survival probability data.
Our findings for all the considered long-baseline experiments are presented in \cref{sec:results}. Finally, we conclude in \cref{sec:conclusion}.

\section{Theory}
\label{sec:Formalism}

Consider a dichotomic observable $\hat{Q}$, whose measurement outputs at any time $t_i$ are restricted to $\pm 1$. The correlation function associated with measurements at times $t_i$ and $t_j$ is defined as
\begin{equation}
    C_{ij} \equiv \langle \hat{Q}(t_i)\,\hat{Q}(t_j) \rangle,
\end{equation}
where $\langle \cdots \rangle$ denotes the expectation value, typically evaluated over many realizations of the experiment \cite{Fritz:2010qzm, Emary:2013wfl}.

Using these correlation functions, the LG parameter, $K_n$, for measurements at $n$ distinct times is defined as
\begin{equation}
    K_n = \sum_{i=1}^{n-1} C_{i\, (i+1)} - C_{n1} \;,
    \label{eq:Kn_standard}
\end{equation}
with the simplest LG measure defined as $K_3 = C_{12} + C_{23} -C_{31}$.
Classically, the measure $K_n$ is bound as~\cite{Emary:2013wfl, Leggett:1985zz}
\begin{equation}
    K_n \leq n-2 \; ,
    \label{eq:Kn_bound}
\end{equation}
for $n \geq 3$. This is known as the Leggett-Garg Inequality (LGI).

In neutrino oscillation experiments, a violation of the LGI can arise either from the intrinsic quantum nature of flavor oscillations or from statistical and systematic uncertainties in realistic measurements. It is therefore necessary to consider the corresponding classical LG measure.
In several previous studies~\cite{Formaggio:2016cuh, Fu:2017hky}, the classical LG measure has been defined as
\begin{equation}
    K_n^C \equiv \sum_{i=1}^{n-1} C_{i\,(i+1)} - \prod_{i=1}^{n-1} C_{i\,(i+1)} \; .
    \label{eq:knclassical}
\end{equation}
This is because, for a time-reversal symmetric two-level classical system, the correlation function $C_{n1}$ can be expressed as
\begin{equation}
    C_{n1} = \prod_{i=1}^{n-1} C_{i\,(i+1)}\; .
\end{equation}
The simplest LG measure (for $n=3$) takes the form $K_3^C = 1 - (1 - C_{12})(1 - C_{23})$. For $0 \leq C_{12}, C_{23} \leq 1$, it can be shown that $K_3^C$ always satisfies the Leggett–Garg inequality $K_3 \leq 1$.
This result can be generalized for $K_n^C$, and thus, classically, no violations of the LGI should be observed. However, for neutrino experiments, the large statistical and systematic uncertainties on the neutrino oscillation probability can lead to values of the correlation function below zero or above 1, which would lead to violations of the LGI.

We also note that a different technique to model the classical behavior was explored in Ref.~\cite{Groth:2025gtf}, and we will discuss this in more detail in~\cref{sec:explicit_strings_quantum_classical}. Our analysis also involves comparing the quantum distribution against such classical predictions.

\subsection{Application at Neutrino Oscillation Experiments}
\label{sec:application_energy}

In long-baseline neutrino experiments, performing measurements at multiple time intervals is impractical, as the distance from the neutrino source to the neutrino detector is fixed. However, special relativity allows us to circumnavigate this issue, as a particle of mass $m$ evolving for a proper time $\Delta\tau$ is associated, in the lab frame, to an interval $\Delta t = (\Delta\tau/m) E$, where $E$ is its energy. Thus, the evolution of the system depends only on the ratio $\Delta t/E$, and varying the neutrino energy at a fixed baseline effectively emulates measurements at different times. For constructing the simplest LG measure $K_3$, we therefore map~\cite{Formaggio:2016cuh, Chattopadhyay:2023xwr}
\begin{equation}
    \frac{\Delta t_{21}}{E_0},\ \frac{\Delta t_{32}}{E_0},\ \frac{\Delta t_{31}}{E_0}
    \quad \longleftrightarrow \quad
    \frac{\Delta t_0}{E_{21}},\ \frac{\Delta t_0}{E_{32}},\ \frac{\Delta t_0}{E_{31}} \; .
    \label{eq:energy_scaling}
\end{equation}
where $\Delta t_{ij} \equiv t_j -t_i$, and $E_{21}$, $E_{32}$, and $E_{31}$ are proxy neutrino energies chosen to satisfy
\begin{equation}
    \frac{1}{E_{21}} + \frac{1}{E_{32}} = \frac{1}{E_{31}} \; .
    \label{eq:energy_relation}
\end{equation}
In what follows, we refer to each viable combination of such 3 energies as an energy triplet. Eq.~(\ref{eq:energy_relation}) arises because the three relevant time intervals must also obey the relation
\begin{equation}
    \Delta t_{21} + \Delta t_{32} = \Delta t_{31}\; .
    \label{eq:time_relation}
\end{equation}
In the context of LGI violation, the above-described mapping allows us to use available neutrino data at various energies in place of measurements that would otherwise require repeated preparation and measurement at different times.
Further, this method depends only on arguments from special relativity and not on the explicit neutrino evolution formula.

\subsection{Generalized Leggett-Garg Strings}

Eq.~(\ref{eq:Kn_standard}) is one possible definition of the LG measure following the usual time ordering.
In general, Leggett–Garg inequalities allow for a broader class of combinations, characterized by arbitrary signs multiplying the correlation functions.
More generally, one can define~\cite{Emary:2013wfl,Halliwell:2015xfa,Blasone:2022iwf}
\begin{equation}
    K_n(\{\sigma_i\}) = \sum_{i=1}^n \sigma_i C_{i\,(i+1)} \; , 
    \label{eq:Kn_general}
\end{equation}
with the identification $C_{n\,(n+1)} \equiv C_{n1}$, and where $\sigma_i = \pm 1$.
The sign assignments must satisfy the condition
\begin{equation}
    \prod_{i=1}^n \sigma_i = -1,
    \label{eq:sign_condition}
\end{equation}
which ensures that the constructed object is a proper LG string.
This generalized approach creates $2^{n-1}$ independent LGIs for $n$ measurements, each of which, under classical realism, must satisfy
\begin{equation}
    |K_n(\{\sigma_i\})| \leq n - 2 \; .
\end{equation}

The standard LG string in~\cref{eq:Kn_standard} corresponds to $\sigma_i = 1$ for $i = 1, \dots, n-1$ and $\sigma_n = -1$.
However, the freedom of selecting the $\{\sigma_i\}$ allows for optimization: for any given set of measured correlation functions, one may define the optimal LG parameter for a test of classicality as the maximum value over all allowed sign assignments:
\begin{equation}
    K_n^{\text{max}} = \max_{\{\sigma_i\}} K_n(\{\sigma_i\}).
\end{equation}
This flexibility can be particularly useful for experimental analyses, as it allows a more robust search for LGI violations by selecting the most sensitive LG string for a given data set~\cite{Groth:2025gtf}.

\subsubsection{Explicit Leggett-Garg Strings for the Quantum and Classical Cases}
\label{sec:explicit_strings_quantum_classical}

For the simplest LG measure $K_3$, we can construct four distinct LG strings for both the quantum and classical scenarios.
The quantum $K_3$ strings are given by
\begin{align}
    K_3^{Q++} &\equiv C_{12} + C_{23} - C_{31} \; ,\nonumber \\
    K_3^{Q+-} &\equiv C_{12} - C_{23} + C_{31} \; ,\nonumber \\
    K_3^{Q-+} &\equiv -C_{12} + C_{23} + C_{31} \; , \nonumber \\
    K_3^{Q--} &\equiv -C_{12} - C_{23} - C_{31}  \; .
    \label{eq:k3quantumstrings}
\end{align}
Here, $C_{ij}$ denotes the correlation function between measurements at times $t_i$ and $t_j$. The superscripts $(++), (+-), (-+),$ and $(--)$ correspond to the sign choices for $\sigma_1$, and $\sigma_2$, with the sign of $\sigma_3$ set by $\sigma_1 \sigma_2 \sigma_3 = -1$.

The corresponding classical expressions for the $K_3$ strings can be systematically constructed from~\cref{eq:k3quantumstrings} by choosing the appropriate sign assignments. Explicitly, these are
\begin{align}
    K_3^{C++} &\equiv C_{12} + C_{23} - C_{12} \cdot C_{23} \nonumber \; , \\
    K_3^{C+-} &\equiv C_{12} - C_{23} + C_{12} \cdot C_{23} \nonumber \; , \\
    K_3^{C-+} &\equiv -C_{12} + C_{23} + C_{12} \cdot C_{23} \nonumber \; ,\\
    K_3^{C--} &\equiv -C_{12} - C_{23} - C_{12} \cdot C_{23} \; ,
    \label{eq:k3classicalstrings}
\end{align}
where $C_{ij}$ now denotes the classical correlation function. As in the quantum case, the superscripts indicate the sign choices $(\sigma_1, \sigma_2)$.
The recipe outlined above for the classical case closely follows Ref.~\cite{Formaggio:2016cuh} and will be referred to as the classical Factorized Correlator (FC) method hereafter.

Another way to model the classical case is to fit the correlators in~\cref{eq:k3quantumstrings} with a function $C_{ij} = e^{-\Gamma L/E_{ij}}$~\cite{Groth:2025gtf} (where $E_{ij}$ corresponds to a particular energy bin), instead of replacing $C_{31} \to C_{12} \, C_{23}$. This will be referred to as the classical Exponential Fit (EF) method hereafter.

In practice, in both the quantum and classical cases, one can obtain the string that yields the strongest violation
\begin{equation}
    K_3^{\max} = \max \Big(K_3^{++}, \, K_3^{+-},\, K_3^{-+},\, K_3^{--}\Big) \; .
    \label{eq:k3max}
\end{equation}
This enhances the sensitivity of LGI-based tests in experiments. Note that, violation of the LGI in only one of the four LG strings would still be a direct hint for the quantum nature of neutrino oscillations; therefore, the quantity $K_3^{\mathrm{max}}$ allows us to capture the deviation from classicality for such systems in a more efficient manner.

\subsubsection{Leggett-Garg Strings in terms of Neutrino Survival Probabilities}

In long-baseline neutrino experiments, the effects of all three neutrino flavors need to be considered.
However, the system can be reduced to two effective flavors: muon neutrinos, $\nu_\mu$, and an effective flavor $\nu_x$ representing the combined state of $\nu_e$ and $\nu_\tau$.
We take the dichotomic observable $\hat{Q}$ to give $+1$ for $\nu_\mu$ and $-1$ for $\nu_x$.
The correlation function $C_{ij}$ is then directly related to the measurable neutrino survival probability $P_{\mu\mu}$ as
\begin{equation}
    C_{ij} = 2P_{\mu\mu}(t_i, t_j) - 1 \; .
    \label{eq:C_survival}
\end{equation}
Here, $P_{\mu\mu}(t_i, t_j)$ is the probability that a muon neutrino is detected at time $t_j$, given it was produced as a muon neutrino at time $t_i$. 
This allows all LG strings to be expressed solely in terms of neutrino survival probabilities.

Substituting~\cref{eq:C_survival} into the expressions of~\cref{eq:k3quantumstrings}, we obtain the explicit forms for the four $K_3$ strings in terms of the neutrino survival probabilities
\begin{align}
    K_3^{Q++} &= -1 + 2P_{21} + 2P_{32} - 2P_{31} \; , \nonumber \\
    K_3^{Q+-} &= -1 + 2P_{21} - 2P_{32} + 2P_{31} \; , \nonumber\\
    K_3^{Q-+} &= -1 - 2P_{21} + 2P_{32} + 2P_{31} \; , \nonumber\\
    K_3^{Q--} &= 3 - 2P_{21} - 2P_{32} - 2P_{31} \; ,
    \label{eq:k3quantumstringsprobability}
\end{align}
where we define $P_{ij} \equiv P_{\mu\mu}(t_i,t_j)$ for compactness.
For the classical FC case, the four $K_3$ strings can be expressed similarly; \cref{eq:k3classicalstrings,eq:C_survival} together give
\begin{align}
    K_3^{C++} &= -3 + 4P_{21} + 4P_{32} - 4P_{21}P_{32} \; , \nonumber \\
    K_3^{C+-} &= 1 - 4P_{32} + 4P_{21}P_{32} \; , \nonumber  \\
    K_3^{C-+} &= 1 - 4P_{21} + 4P_{21}P_{32} \; , \nonumber  \\
    K_3^{C--} &= 1 - 4P_{21}P_{32} \; .
    \label{eq:k3classicalstringsprobability}
\end{align}
The replacement $P_{31} \to P_{21} P_{32} + (1-P_{21}) (1-P_{32})$ in~\cref{eq:k3quantumstringsprobability} also leads to the set of equations in~\cref{eq:k3classicalstringsprobability}.

To estimate the LG strings using the classical Exponential Fit (EF) method (as employed in~\cite{Groth:2025gtf}), we fit the correlator to the form $C_{ij} = e^{-\Gamma L/E_{ij}}$.
This does not involve explicit expressions for the LG strings in terms of survival probabilities. Instead, we compute the correlators from $C_{ij} = 2P_{ij} - 1$, taking their associated uncertainties from the muon neutrino survival probability data, and determine the best-fit value of $\Gamma$.

By measuring the neutrino survival probability at various energies, one can construct both quantum and classical Leggett–Garg strings directly from experimental data.
The comparison of these quantities therefore provides a stringent test of violations of classical macrorealism in the neutrino sector, and allows us to quantify the degree of quantumness in a given neutrino experiment.

\subsubsection{Comparing $K_3^{Q++}$ and $K_3^{\mathrm{max}}$}

The LG measure $K_3^{Q++}$, often used to quantify deviations from classical behavior, provides a standard test of quantumness (e.g., see~\cite{Formaggio:2016cuh, Fu:2017hky}).
However, $K_3^{\mathrm{max}}$ is a more well-suited alternative; it is defined as the maximum over the four LG strings, and therefore captures the largest possible violation of classicality, if there is any.

\begin{figure}[t!]
    \centering
    \begin{minipage}[c]{0.45\linewidth}
        \centering
        \includegraphics[width=\linewidth]{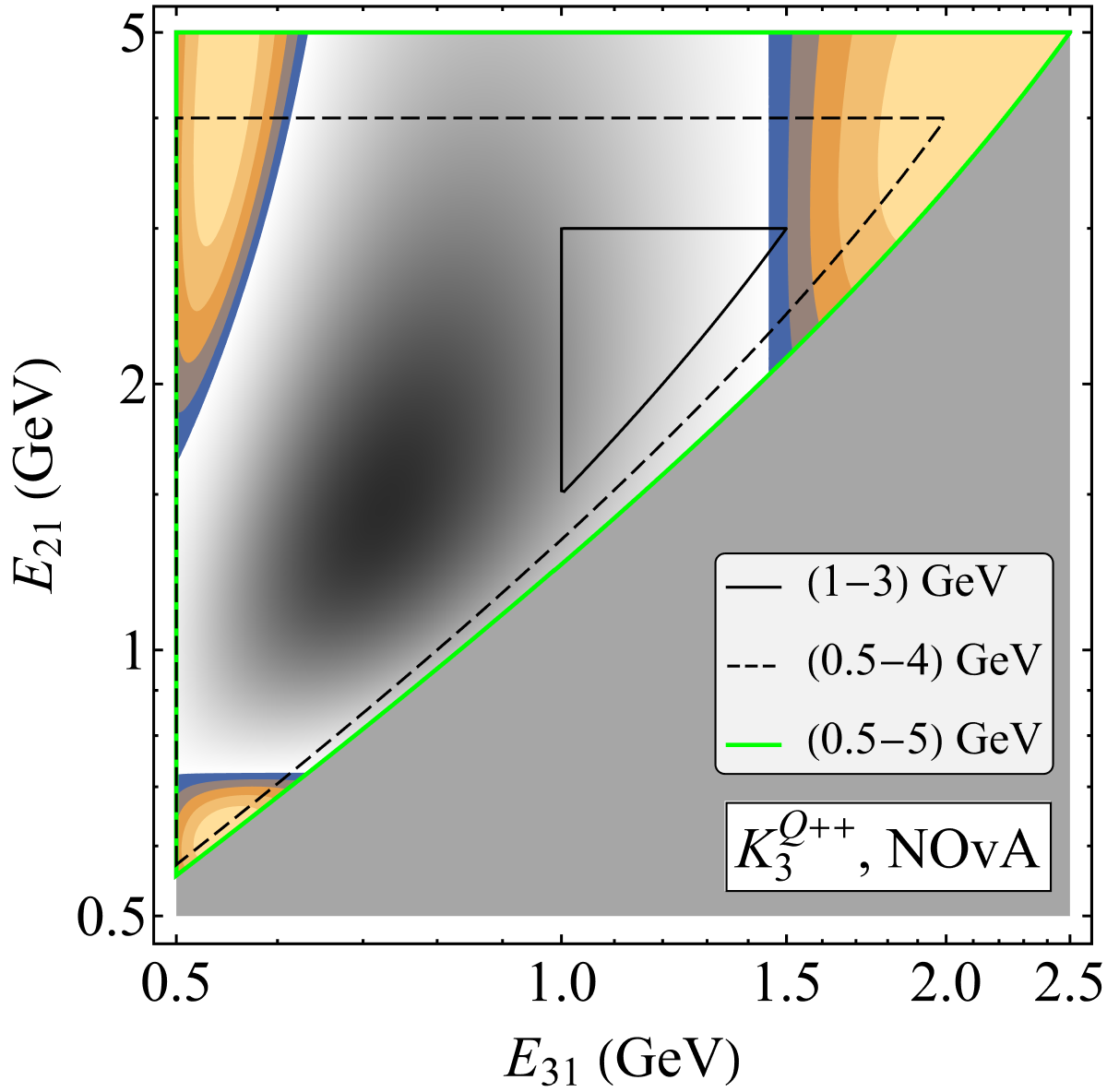}
    \end{minipage}
    \hfill
    \begin{minipage}[c]{0.45\linewidth}
        \centering
        \includegraphics[width=\linewidth]{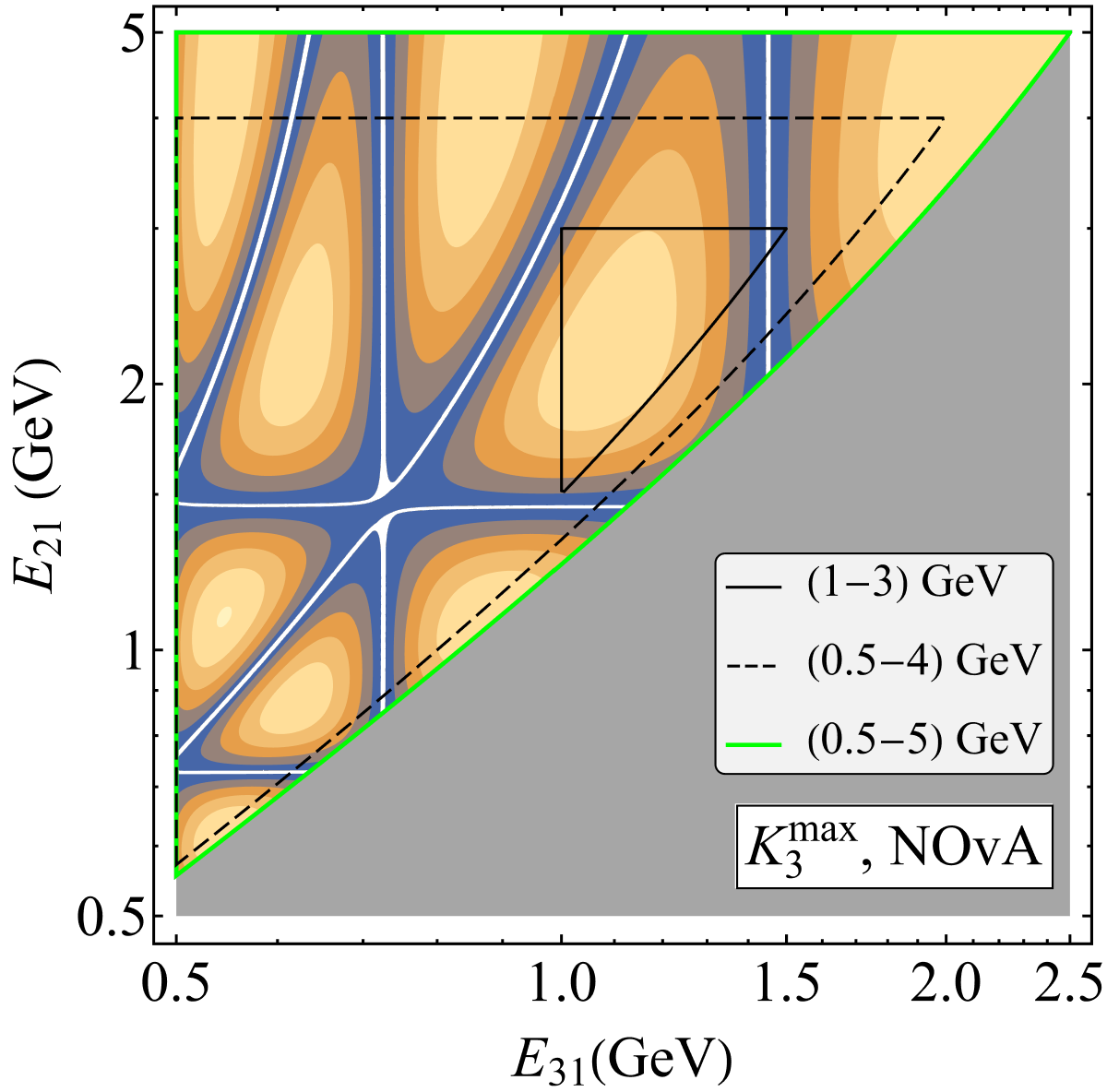}
    \end{minipage}
    \hfill
    \begin{minipage}[c]{0.06\linewidth}
        \centering
        \raisebox{0.8cm}{\includegraphics[width=\linewidth]{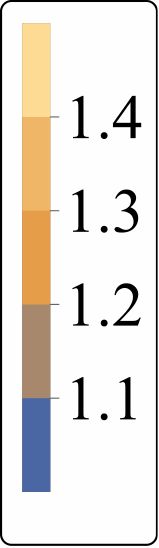}}
    \end{minipage}
    \caption{
$K_3^{Q++}$ (left panel) and $K_3^{\mathrm{max}}$ (right panel) for the neutrino energies accessible at the NOvA experiment. The axes correspond to neutrino energies $E_{31}$ and $E_{21}$ (see~\cref{sec:application_energy}), with the shaded region denoting the disallowed values of $E_{32}$ (corresponding to invalid energy triplet choices). The colorful regions denote $K_3 > 1$ values (values up to $K_3 \lesssim 1.5$ are presented, as indicated in the legend), where the nonclassical behavior of the NOvA experiment may be expected.
    }
    \label{fig:NOvA_K3_comparison}
\end{figure}

In~\cref{fig:NOvA_K3_comparison}, we compare these two different quantities for the quantum scenario. We show the quantity $K_3^{Q++}$ in the left panel, and $K_3^{\mathrm{max}}$ in the right panel. The colourful regions denote values of $K_3 > 1$, with the different colors corresponding to values of $K_3$ up to $\sim1.5$. We use the NOvA experiment for this comparison, and employ benchmark values for the neutrino mixing parameters consistent with the global fit~\cite{deSalas:2020pgw,Capozzi:2021fjo,Esteban:2024eli}:
\begin{align}
    \Delta m^2_{31} = 2.5 \times 10^{-3}~\text{eV}^2\; ,& \quad \Delta m^2_{21} = 7.5 \times 10^{-5}~\text{eV}^2 , \nonumber \\
    \theta_{12} = 33.5^\circ\;, \quad \theta_{23} = 45^\circ \;, & \quad \theta_{13} = 8.5^\circ \; , \quad \delta_{\text{CP}} = -90^\circ \; .
    \label{eq:oscparam}
\end{align}
The gray triangular region outside the green boundaries is unphysical since, for NOvA, no energy triplet exists in that region. The black and white region in the left panel satisfies the LGI of $K_3 <1$.

Note that, while the LG measure $K_3^{Q++}$ is able to capture the quantum behavior only for certain regions of the parameter space, the quantity $K_3^{\mathrm{max}}$ captures the violation of the LGI for all available neutrino energies. This conclusion remains valid regardless of the precise choice of neutrino mixing parameters and holds across neutrino experiments, since the quantity $K_3^{\mathrm{max}}$ can capture a violation of the LGI as long as at least one of the four measures, $K_3^{Q++}$, $K_3^{Q+-}$, $K_3^{Q-+}$, or $K_3^{Q--}$, violates the inequality. This demonstrates that considering all possible LG strings through $K_3^{\mathrm{max}}$ can, in principle, significantly increase the likelihood for observation of non-classicality, further motivating the use of this approach in our analyses.

However, in a realistic experiment, the large error bars in the neutrino survival probability can lead to non-trivial complications. 
Even for the classical case, apparent violations of the LGI may be observed due to statistical and systematic fluctuations in the values of the neutrino survival probability or the correlation function.
Since the relationship $K_3^{\mathrm{max}} \geq K_3^{++}$ always holds, even a purely classical implementation can yield a larger apparent violation of LGI when using $K_3^{\mathrm{max}}$.
In fact, this can, in principle, even lead to an overall reduction in the significance of quantumness for such systems.

Therefore, when using $K_3^{\mathrm{max}}$, a careful treatment of experimental uncertainties is necessary.
A simple way to quantify the degree of quantumness is to count the number of energy triplets that violate the classical bound of $K_3 \leq 1$ in the quantum scenario and compare this with the two classical baselines: the Factorized Correlator (FC) and the Exponential Fit (EF) methods.
Furthermore, to enable comparisons across neutrino experiments, one may consider the fraction of energy triplets (quantum and classical) that violate the LGI.
This allows for a direct comparison between experiments with different numbers of accessible energy triplet combinations, where for each energy triplet one would obtain a corresponding value of $K_3^{\mathrm{max}}$.

\subsection{The RMS \textit{z}-score}
\label{sec:zscore}

Although the fraction of triplets that violate the LGI provides a direct qualitative indication that neutrino oscillations are non-classical, it has two important limitations:
\begin{enumerate}
    \item This quantity is insensitive to the magnitude of violation, i.e., by what amount the LG measure $K_3$ for a particular triplet exceeds the classical bound. As a result, it carries no information about the statistical significance of the observed violation.
    \item In practice, the sparse and non-Gaussian character of the classical distributions is often observed when presenting the fraction of LGI violations. This makes it difficult to quantify the separation between classical and quantum histograms. Hence, it is not straightforward to present a meaningful measure for calculating the significance of classicality violation.
\end{enumerate}
Therefore, in the context of quantumness at neutrino oscillation experiments, Ref.~\cite{Groth:2025gtf} introduced a different metric that assigns weights according to the magnitude of the LGI violations, called the RMS $z$-score.
In~\cref{sec:z_rms}, we will introduce a revised version of the RMS $z$-score and explain why such a modification is suitable for LGI tests at neutrino oscillation experiments.

\subsubsection{Defining the \textit{z}-score}
\label{sec:z_score_defn}

First, for each energy triplet, we define four $z$-scores, corresponding to the four LG strings as
\begin{align}
    z^{++} =  \frac{K_{3}^{++} - 1}{\delta K_{3}^{++}} \; ,& \qquad
    z^{+-} =  \frac{K_{3}^{+-} - 1}{\delta K_{3}^{+-}} \; ,\nonumber\\
    z^{-+} =  \frac{K_{3}^{-+} - 1}{\delta K_{3}^{-+}} \; ,& \qquad
    z^{--} =  \frac{K_{3}^{--} - 1}{\delta K_{3}^{--}} \; .
    \label{eq:z_strings}
\end{align}
Here, the $\delta K_{3}$ denotes the uncertainty in each of the $K_3$ LG strings, and can be calculated by propagating the error from the experimental survival probability data.
In the quantum case (as well as the classical EF case), each energy triplet $(E_{21},\, E_{32},\, E_{31})$ involves three sampled probabilities $(P_{21},\, P_{32},\, P_{31})$. 
For the quantum scenario, the explicit expressions are
\begin{align}
\delta K_{3}^{Q++} = &\,
2\sqrt{(\delta P_{21}^{u})^2 + (\delta P_{32}^{u})^2
      + (\delta P_{31}^{l})^2} \; , \nonumber \\
\delta K_{3}^{Q+-} = &\,
2\sqrt{(\delta P_{21}^{u})^2 + (\delta P_{32}^{l})^2
      + (\delta P_{31}^{u})^2} \; , \nonumber \\
\delta K_{3}^{Q-+} = &\,
2\sqrt{(\delta P_{21}^{l})^2 + (\delta P_{32}^{u})^2
      + (\delta P_{31}^{u})^2} \; , \nonumber \\
\delta K_{3}^{Q--} = &\,
2 \sqrt{(\delta P_{21}^{l})^2 + (\delta P_{32}^{l})^2
      + (\delta P_{31}^{l})^2} \; .
\end{align}
Here, the notation $\delta P_{ij}^{u}$ ($\delta P_{ij}^{l}$) denotes the upper (lower) uncertainty on $P_{ij}$, allowing for the possibility of asymmetric errors in the measured survival probability. Note that, in the above equations, the error bars correspond to fluctuations in the direction which maximizes the LG strings, and thus allow for a larger value of $\delta K_{3}^{Q}$ strings, i.e., the errors are propagated in the direction that increases the value of the LG string (consistent with obtaining an upper bound on their values). Let us explain this explicitly: the first LG string is given by $K_{3}^{Q++} = -1 + 2 P_{21} + 2 P_{32} - 2 P_{31}$, therefore, $K_{3}^{Q++}$ will be larger for larger values of $P_{21}$ and $P_{32}$, and smaller value of $P_{31}$, which is why we define $\delta K_{3}^{Q++}$ with $(\delta P_{21}^{u}, \, \delta P_{32}^{u},\, \delta P_{31}^{l})$.

For obtaining the $z$-scores for the classical EF case, we use the three classical correlators $(C_{12},\, C_{23},\, C_{31})$, with their upper/lower uncertainties propagated similar to the quantum case
\begin{align}
\left[ \delta K_{3}^{C++} \right]_{\rm EF}= & \,
\sqrt{(\delta C_{12}^{u})^2 + (\delta C_{23}^{u})^2
      + (\delta C_{31}^{l})^2} \; , \nonumber \\
\left[\delta K_{3}^{C+-} \right]_{\rm EF}= & \,
\sqrt{(\delta C_{12}^{u})^2 + (\delta C_{23}^{l})^2
      + (\delta C_{31}^{u})^2} \; , \nonumber \\
\left[\delta K_{3}^{C-+} \right]_{\rm EF}= & \,
\sqrt{(\delta C_{12}^{l})^2 + (\delta C_{23}^{u})^2
      + (\delta C_{31}^{u})^2} \; , \nonumber \\
\left[\delta K_{3}^{C--} \right]_{\rm EF}= & \,
\sqrt{(\delta C_{12}^{l})^2 + (\delta C_{23}^{l})^2
      + (\delta C_{31}^{l})^2} \; .
\end{align}
Here, the errors corresponding to the three correlators are propagated from the errors in the survival probability using~\cref{eq:C_survival}; for example, $\delta C_{12}^{u,l} = 2\, \delta P_{21}^{u,l}$.

Finally, for the classical FC baseline, both the LG strings and their associated uncertainties depend only on $P_{21}$ and $P_{32}$.
Due to the structure of the expressions in~\cref{eq:k3classicalstringsprobability}, it is nontrivial to answer whether increasing or decreasing $P_{21}$ or $P_{32}$ values leads to a larger value of the LG strings. To obtain a conservative estimate of the $z$-scores for the classical FC baseline, i.e., to avoid understating the degree of apparent violation for the classical case, we minimize the denominator of~\cref{eq:z_strings} by intentionally underestimating the uncertainties $\delta K_3$. Therefore, we define the errors as
\begin{align}
\left[\delta K_{3}^{C++} \right]_{\rm FC}= &\, 4
\sqrt{(1-P_{32})^2 (\delta P_{21}^{\min})^2 
     + (1- P_{21})^2 (\delta P_{32}^{\min})^2} \; , \nonumber \\
\left[\delta K_{3}^{C+-} \right]_{\rm FC}=&\, 4 
\sqrt{(P_{32})^2 (\delta P_{21}^{\min})^2 
     + (P_{21}-1)^2 (\delta P_{32}^{\min})^2} \; , \nonumber \\
\left[\delta K_{3}^{C-+} \right]_{\rm FC}=&\,4 
\sqrt{(P_{32}-1)^2 (\delta P_{21}^{\min})^2 
     + (P_{21})^2 (\delta P_{32}^{\min})^2} \; , \nonumber \\
\left[\delta K_{3}^{C--} \right]_{\rm FC}=&\, 4
\sqrt{(P_{32})^2 (\delta P_{21}^{\min})^2 
     + (P_{21})^2 (\delta  P_{32}^{\min})^2} \; .
\end{align}
Here, $\delta P_{ij}^{\min} \equiv \min \!\left(\delta P_{ij}^{u},\, \delta P_{ij}^{l}\right)$ denotes the smaller of the upper and lower uncertainties.

The four $z$-scores ($z^{++},\, z^{+-}, \, z^{-+},\,z^{--}$) allow us to quantify the degree of LG violation for each energy triplet for a given experiment, normalized with respect to the intrinsic uncertainty, for each of the four realizations of the $K_3$ LG string. We define the final $z$-score as the maximum between these four quantities and zero,
\begin{equation}
    z = \max \left(z^{++},\, z^{+-},\, z^{-+},\, z^{--} ,\, 0\right)\;.
\end{equation}
This definition ensures that $z \geq 0$ even in the absence of any LG violation (i.e., the lowest possible value of $z$-score is zero).

\subsubsection{Defining the RMS $z$-score}
\label{sec:z_rms}

We define the aggregate $z$-RMS statistic as
\begin{equation}
    z_{\rm RMS} = \sqrt{ \frac{1}{N} \sum_{i=1}^N z_i^2 }\; ,
    \label{eq:zrms}
\end{equation}
with $N$ being the total number of energy triplets. Note that our definition differs from that in Ref.~\cite{Groth:2025gtf}. In Ref.~\cite{Groth:2025gtf}, the RMS statistic is normalized by $N_{\rm LGV}$, which is the number of energy triplets that violate the LGI.

However, normalizing with respect to $N_{\rm LGV}$ can lead to a significant overestimation of the $z_{\rm RMS}$ values, particularly for scenarios with $N_{\rm LGV} \ll N$, where $N$ is the total number of triplets. For example, let's take 2  experiments with a similar number of total triplets, $\mathcal{O}(N^{(1)}) \approx \mathcal{O}(N^{(2)})$. Further, we take that  $\mathcal{O}(N^{(1)}_{\rm LGV}) \ll \mathcal{O}(N^{(2)}_{\rm LGV})$. If the $z$-score values for the LG violating triplets are comparable across the two experiments, then normalizing with $N_{\rm LGV}$ would lead to comparable $z_{\rm RMS}$ values for both experiments. Essentially, this is the opposite problem of simply looking at the fraction of triplets that violate the LGI.
In that case, all violations are weighted equally, regardless of their magnitude. In contrast, if one normalizes the RMS z-score by $N_{\rm LGV}$, the measure becomes insensitive to how many triplets violate the LGI in the first place.
However, if we normalize with respect to the total number of triplets $N$, then the $z_{\rm RMS}$ value for experiment $(1)$, where only a few triplets violate the LGI, would drop significantly.
Therefore, in our definition, the $z_{\rm RMS}$ is normalized by $N$, the total number of triplets, and this allows our measure to be sensitive to both the magnitude and frequency of violation of the LGI.

To quantify the degree of deviation from classical behavior, we have to implement a systematic, step-by-step analysis. In what follows, we cover the key details necessary for such an approach.

\section{Implementation}
\label{sec:method}

In this section, we address some of the key technical challenges, including the effect of asymmetric uncertainties in the neutrino survival probability data and the identification of suitable energy triplets for calculating the LG strings.
The framework outlined allows for a direct statistical comparison between quantum and classical scenarios.

\subsection{Obtaining the Energy Triplets for the Leggett-Garg Strings}
\label{sec:energy_triplets}

One key step in our study is the systematic determination of energy triplets for testing the LGI. We label the three energies of the triplet as $E_{21}$, $E_{32}$, and  $E_{31}$. Each triplet corresponds to three energy ranges (also referred to as bins) which can be specified via their left and right edges, i.e. $E_{ij} \equiv (E_{ij}^L,\, E_{ij}^R)$. To construct all possible combinations of the energy triplet for an experiment, we first take two of the energy bins directly from  experimental data ($E_{21}$ and $E_{32}$), and then construct the third energy range, $E_{31}$, such that
\begin{equation}
    \widetilde{E}_{31}^{L,\, R} = \left( \frac{1}{E_{21}^{L,\, R}} + \frac{1}{E_{32}^{L,\, R}} \right)^{-1}
    \label{eq:harmonic_overlap}
\end{equation}
is satisfied for both left and right edges. The $\widetilde{E}_{31}$ (tilde) denotes that this energy bin is generated from $E_{21}$ and $E_{32}$.

The resultant $(\widetilde{E}_{31}^L,\, \widetilde{E}_{31}^R)$ is typically not going to correlate exactly with any experimental bin, even when $E_{21}$ and $E_{32}$ are obtained directly from the available experimental energy bins. We refer to $\widetilde{E}_{31}$ as the ``generated'' bin in what follows.

\begin{figure}[t]
    \centering
    \includegraphics[width=0.9\textwidth]{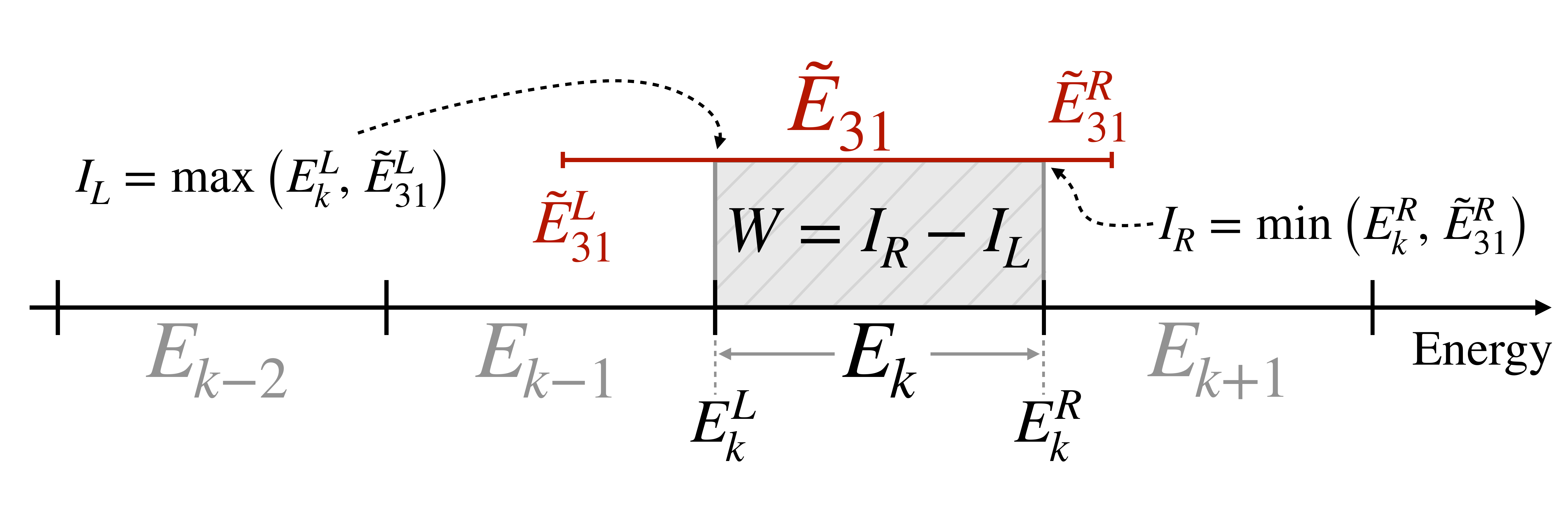}
    \caption{Schematic diagram representing the overlap between the generated energy bin $(\widetilde E_{31}^L, \widetilde E_{31}^R)$ and $k^{th}$ experimental energy bin $(E_{k}^L, E_{k}^R)$. The gray shaded area corresponds to the overlapping region where its left and right boundaries are defined by $I_L =\max(E_{k}^L, \, \widetilde E_{31}^L) $ and $I_R =\min(E_{k}^R, \, \widetilde E_{31}^R) $, respectively.}
    \label{fig:energy_bin_intersection}
\end{figure}

Next, we perform a detailed analysis to find the experimental energy bin that maximally overlaps with $\widetilde{E}_{31}$. We first compute the width of the overlap of the generated interval $(\widetilde{E}_{31}^L, \, \widetilde{E}_{31}^R)$ with the $k$-th experimental bin $(E_{k}^L,\, E_{k}^R)$, by defining:
\begin{align}
    \text{Left edge of the overlap:} \qquad & I_L = \max(E_{k}^L, \, \widetilde E_{31}^L)  \; , \nonumber\\
    \text{Right edge of the overlap:} \qquad & I_R= \min(E_{k}^R, \, \widetilde E_{31}^R) \; , \nonumber \\
    \text{Width of overlap:} \qquad & W = \max(0, \, I_R - I_L) \; .
\end{align}
This is shown schematically in~\cref{fig:energy_bin_intersection}. For a scenario where the generated energy bin ($\widetilde E_{31}$) overlaps with multiple experimental energy bins, we can consider two different ways to measure the amount of relative overlap.
In the first method, we divide the width of the overlap ($W$) by the width of the generated $\widetilde E_{31}$ energy bin, i.e., we express the overlap relative to the generated energy bin. In the second method, we divide the width of the overlap by the width of the experimental energy bin ($E_{k}$).
We express the two definitions as follows:
\begin{align}
    O_{\rm gen} &= \frac{W}{\widetilde E_{31}^R - \widetilde E_{31}^L}\; , \nonumber \\
    O_{\rm exp} &= \frac{W}{E_{k}^R - E_{k}^L}\; .
\end{align}
Here, $\widetilde E_{31}^L$ and $\widetilde E_{31}^R$ denote the left and right edges of the generated energy bin, while $E_{k}^L$ and $E_{k}^R$ correspond to the left and right edges of the $k^{\text{th}}$ experimental energy bin.
\begin{figure}[t]
    \centering
    \includegraphics[width=0.8\textwidth]{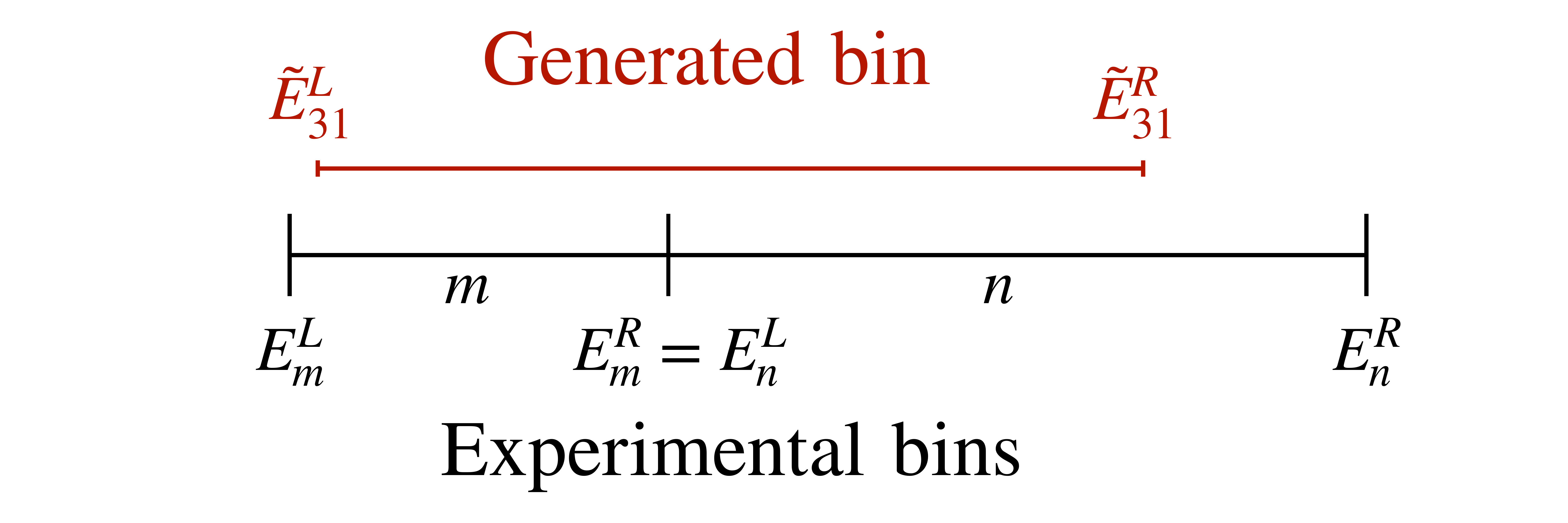}
    \caption{Schematic diagram representing the overlap between the generated energy bin $(\widetilde E_{31}^L, \widetilde E_{31}^R)$ and two experimental energy bins ($m$ and $n$).}
    \label{fig:overlap}
\end{figure}

The overlap of the generated energy bin with multiple experimental energy bins can lead to non-trivial complications, especially if the width of the experimental energy bins is not uniform. We point out one such scenario in~\cref{fig:overlap} by taking a simple example where one generated energy bin overlaps with two experimental energy bins (denoted as $m$ and $n$) of unequal width. In the scenario shown in the figure, $O_{\rm gen}(m) < O_{\rm gen}(n)$ is satisfied, since energy bin $n$ covers a larger fraction of the $(\widetilde E_{31}^L,\, \widetilde E_{31}^R)$ range. However, a bigger fraction of the energy bin $m$ is covered by the range $(\widetilde E_{31}^L,\, \widetilde E_{31}^R)$ in our example, and thus $O_{\rm exp}(m) > O_{\rm exp}(n)$. Therefore, which experimental bin represents best the theoretically generated estimate depends on how we define the overlap.
We define the overlap score as the product 
\begin{equation}
    {\rm overlap~score} = O_{\rm gen} \times O_{\rm exp} \; ,
\end{equation}
so that both arguments are considered in our final selection.

Among all candidate experimental bins considered in the analysis, we select the one with the highest overlap score, identifying it as the best representative of the generated $(\widetilde E_{31}^L, \, \widetilde E_{31}^R)$ interval.
For every combination of $E_{21}$ and $E_{32}$, we first estimate $\widetilde{E}_{31}$ in the dataset, and then calculate the overlap score, creating a complete set of viable triplets for each long-baseline neutrino experiment.
 Our search strategy ensures that for each combination of $E_{21}$ and $E_{32}$, we find only one suitable $E_{31}$, i.e., there is no double counting.
The procedure devised in this section ensures that the chosen energy triplets are physically well-motivated.

\subsection{Sampling in the Presence of Asymmetric Probability Uncertainties}
\label{sec:sampling}

To calculate the LG measures and the RMS $z$-score, first, we need to assign an oscillation probability (or correlator) value corresponding to an energy bin.
To do this, we sample the probability around its mean value and properly deal with the corresponding uncertainty ranges.
In many cases, experimental oscillation probability data have asymmetric uncertainties, meaning that the upper and lower error bars have different lengths.
To incorporate the effects of these error bars, we introduce a split-Gaussian Monte Carlo (MC) approach to precisely map these uncertainties into our statistical sampling.

\begin{figure}[t]
    \centering
    \includegraphics[width=0.5\textwidth]{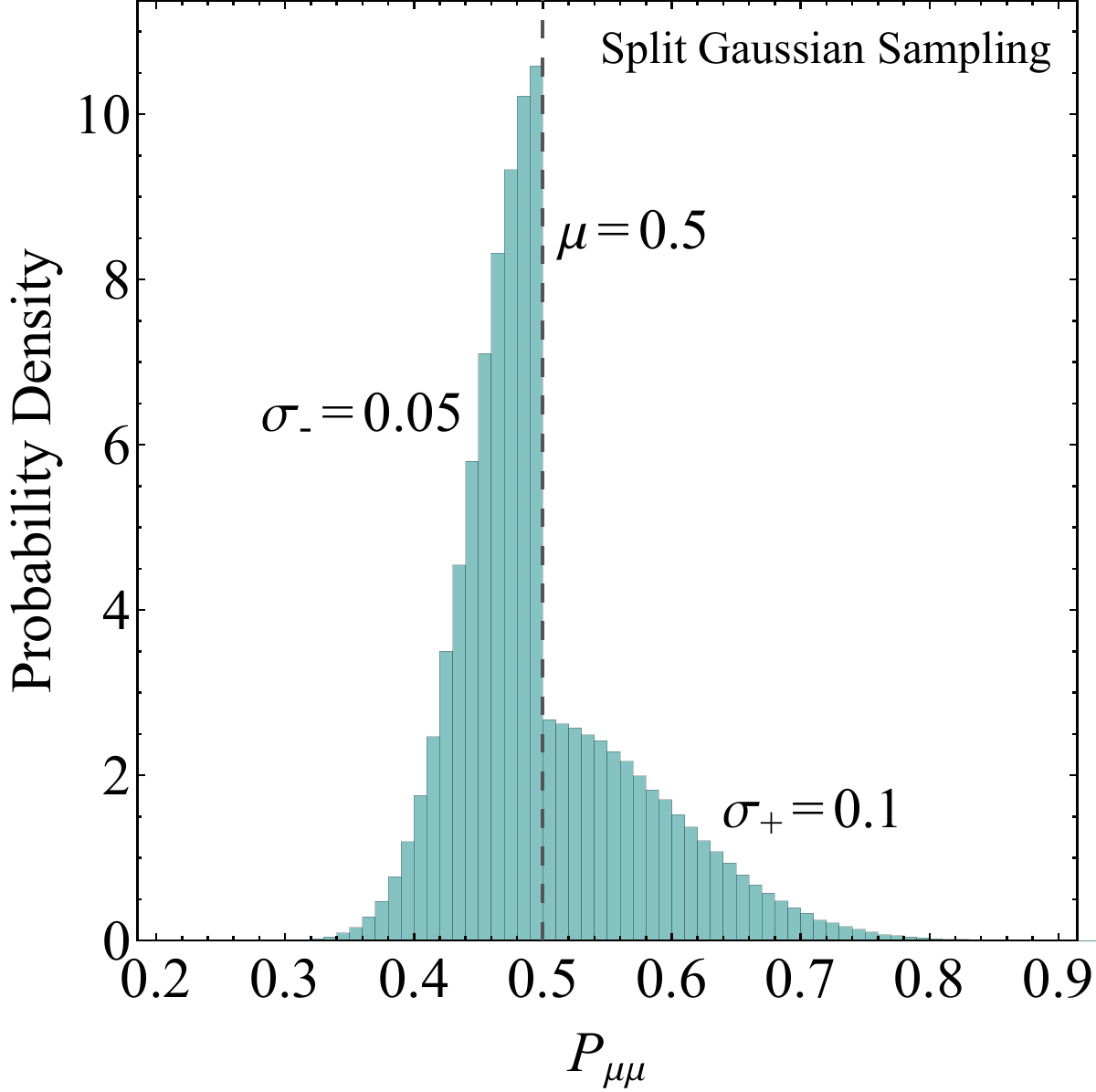}
    \caption{Probability density function obtained through a split-Gaussian sampling of an asymmetric mock oscillation probability distribution, with mean $\mu = 0.5$, and upper and lower standard deviations given by $\sigma_+ = 0.1$, and $\sigma_- = 0.05$, for $10^6$ MC samples.}
    \label{fig:splitgaussian}
\end{figure}

We create MC samples for each observable with a measured central value $\mu$ with asymmetric uncertainties $\sigma_{-}$ (lower) and $\sigma_{+}$ (upper), by following the steps given below:
\begin{enumerate}
    \item Generate a uniform random variable $u \in [0, 1]$.
    \item Compute $w = \sigma_{+}/(\sigma_{-} + \sigma_{+})$.
    \item If $u < w$, take a sample from the lower Gaussian half
    \[
        x = \mu - |N(0, \sigma_{-})| \; ,
    \]
    otherwise (for $u \geq w$), take a sample from the upper Gaussian half
    \[
        x = \mu + |N(0, \sigma_{+})| \; .
    \]
\end{enumerate}
Here, $N(0, \sigma)$ denotes a random variate drawn from a normal distribution with mean $0$ and standard deviation $\sigma$. The absolute value ensures that the fluctuation is oriented away from the center value, and the sign in front ensures that the fluctuation is in the correct direction. In~\cref{fig:splitgaussian}, we show the results of the split-Gaussian sampling technique given above, for a mean of $\mu=0.5$, and $\sigma_+ = 0.1$, $\sigma_-=0.05$, where we have taken $10^6$ MC samples to build the probability density function.

This approach preserves both the central value and the observed asymmetry of the experimental uncertainties. The split-Gaussian sampling is necessary to accurately determine the fraction of triplets that violate LGI, as well as any other observable whose interpretation depends on treating the underlying uncertainties properly.

\section{Results}
\label{sec:results}

In this section, we present the statistical significance of LGI violations for four long-baseline neutrino experiments: MINOS, NOvA, T2K, and DUNE. These neutrino experiments have varying baselines and different neutrino energy ranges, providing complementary probes of the quantum nature of neutrino oscillations.
To quantify the quantumness for each experiment, we will be comparing the quantum and classical predictions using $(i)$ the distributions of the fraction of energy triplets that violate the LGI and $(ii)$ the distributions of the RMS $z$-score.
We construct the relevant distributions by sampling muon neutrino survival probabilities in each energy bin (with mean and standard deviation taken from experimental data), for all energy triplets in a given experiment, and computing both $N_{\rm LGV}$ and $z_{\rm RMS}$ for each sample. This procedure is repeated $10^{5}$ times to obtain accurate representations of the distributions corresponding to quantum and classical behavior.

The quantum scenario follows the procedure outlined in~\cref{eq:k3quantumstrings,eq:k3quantumstringsprobability}. 
For the classical case, to account for apparent LGI violations that may arise from statistical and systematic uncertainties in the data, we consider two approaches:
\begin{enumerate}
    \item Classical Factorized Correlator (FC) method: We impose the classical relation $C_{31} = C_{21} \cdot C_{32}$, and express all correlators in terms of survival probabilities, see~\cref{eq:k3classicalstringsprobability} and detailed discussion in~\cref{sec:Formalism}. This method was used in~\cite{Formaggio:2016cuh}.
    
    \item Classical Exponential Fit (EF) method: Here, the correlators are modeled as $C_{ij} = \exp(-\Gamma L/E_{ij})$, with the parameter $\Gamma$ determined from the fit to the data. The uncertainties on $C_{ij}$ are obtained by propagating the experimental errors on the survival probabilities for each considered experiment. This method was employed in~\cite{Groth:2025gtf}.
\end{enumerate}

Before evaluating the significance of LGI violations for each experiment, we briefly summarize the relevant details of the four experiments considered in this work, including the references used for the probability-level data (with associated uncertainties).

\smallskip \smallskip
\noindent
MINOS: The Main Injector Neutrino Oscillation Search (MINOS) experiment, located in the US, used a 735 km baseline from Fermilab in Illinois to the Soudan mine in Minnesota. It primarily measured muon-neutrino disappearance with beam energies in the $1–10$ GeV range, with the flux peaking at $\sim 3$ GeV. For the muon neutrino survival probability data used in our analysis, we employ the MINOS results from Ref.~\cite{Sousa:2015bxa, Holin:2015lya}.

\smallskip \smallskip
\noindent
NOvA: The NuMI Off-Axis $\nu_e$ Appearance (NOvA) experiment, also in the US, has an 810 km baseline from Fermilab to Ash River, Minnesota. Its off-axis design produces neutrinos mostly between 1 and 4 GeV, with the flux peaked near 2 GeV. In this work, we use the muon neutrino survival probability data from Ref.~\cite{Catano-Mur:2022kyq}.

\smallskip \smallskip
\noindent
T2K: The Tokai to Kamioka (T2K) experiment in Japan sends a neutrino beam from J-PARC in Tokai to the Super-Kamiokande detector in Kamioka, 295 km away. Its off-axis design yields a narrow energy spectrum in the $\sim (0.2–2)$ GeV energy range, peaked near 0.6 GeV. For T2K, we take the muon neutrino survival probability data from Ref.~\cite{patrick_dunne_2020_3959558}. This data has not been previously analyzed in the context of quantum behavior at neutrino experiments.

\smallskip \smallskip
\noindent
DUNE: The Deep Underground Neutrino Experiment (DUNE), currently under construction in the US, will employ a 1300 km baseline from Fermilab to the Sanford Underground Research Facility (SURF) in South Dakota. This will be the longest baseline for oscillation studies using neutrinos produced in accelerators. DUNE is designed to cover an energy range of $\sim(0.5–8)$ GeV, with the beam flux peaking around $2.5$ GeV. In the absence of any recent and publicly available pseudodata, we construct the mock data as follows.

For DUNE, in the energy range of $(0.5\,–\,7.75)$ GeV, we take bin sizes of $0.25$ GeV (leading to 29 energy bins), and for each energy bin, we assign the statistical error in the oscillation probability from Fig.~10 of Ref.~\cite{DUNE:2020jqi}. Note that the data in Ref.~\cite{DUNE:2020jqi} features a relative error of $1/\sqrt{N_{E_\nu}}$, where $N_{E_\nu}$ is the number of events in the energy bin $E_\nu$. Therefore, the statistical error at the probability level can be simply expressed as:
\begin{equation}
    \delta P_{\rm stat}(E_\nu) = P_{\mu\mu}(E_\nu)/ \sqrt{N_{E_\nu}} \; .
    \label{eq:stat_error}
\end{equation}
For every energy bin, we compute the oscillation probability by integrating the full three-flavor muon neutrino survival probability over the width of the energy bin and dividing by the bin width. We take a baseline of $1300$ km and the oscillation parameters from~\cref{eq:oscparam}.
We include systematic errors, taking a benchmark choice of 5\% systematic error at the probability level (for the treatment with an array of possible systematic errors, see~\cref{sec:varying_systematic}), i.e.,
\begin{equation}
    \delta P_{\rm syst}(E_\nu) = 0.05 \; .
    \label{eq:benchamrk_syst_error}
\end{equation}
Therefore, the total error in each individual energy bin is given by
\begin{equation}
    \delta P = \sqrt{\left(\delta P_{\rm syst}\right)^2+\left(\delta P_{\rm stat}\right)^2} \; .
    \label{eq:tot_error}
\end{equation}
This simplified treatment of errors differs from a realistic experimental analysis in two aspects: $(i)$ the systematic error is an energy-dependent quantity, and $(ii)$ the error obtained in our analysis is symmetric, with both upper and lower error bars in generated pseudodata being of the same length. Nevertheless, the DUNE results presented in this section hold as a proof-of-concept, and can be updated in the future with actual data. The DUNE results shown in this work correspond to a benchmark with a 5\% systematic error, as considered in~\cite{DUNE:2015lol}.

\begin{figure}[t]
    \centering
    \begin{minipage}{0.48\textwidth}
        \centering
           \includegraphics[width=\linewidth]{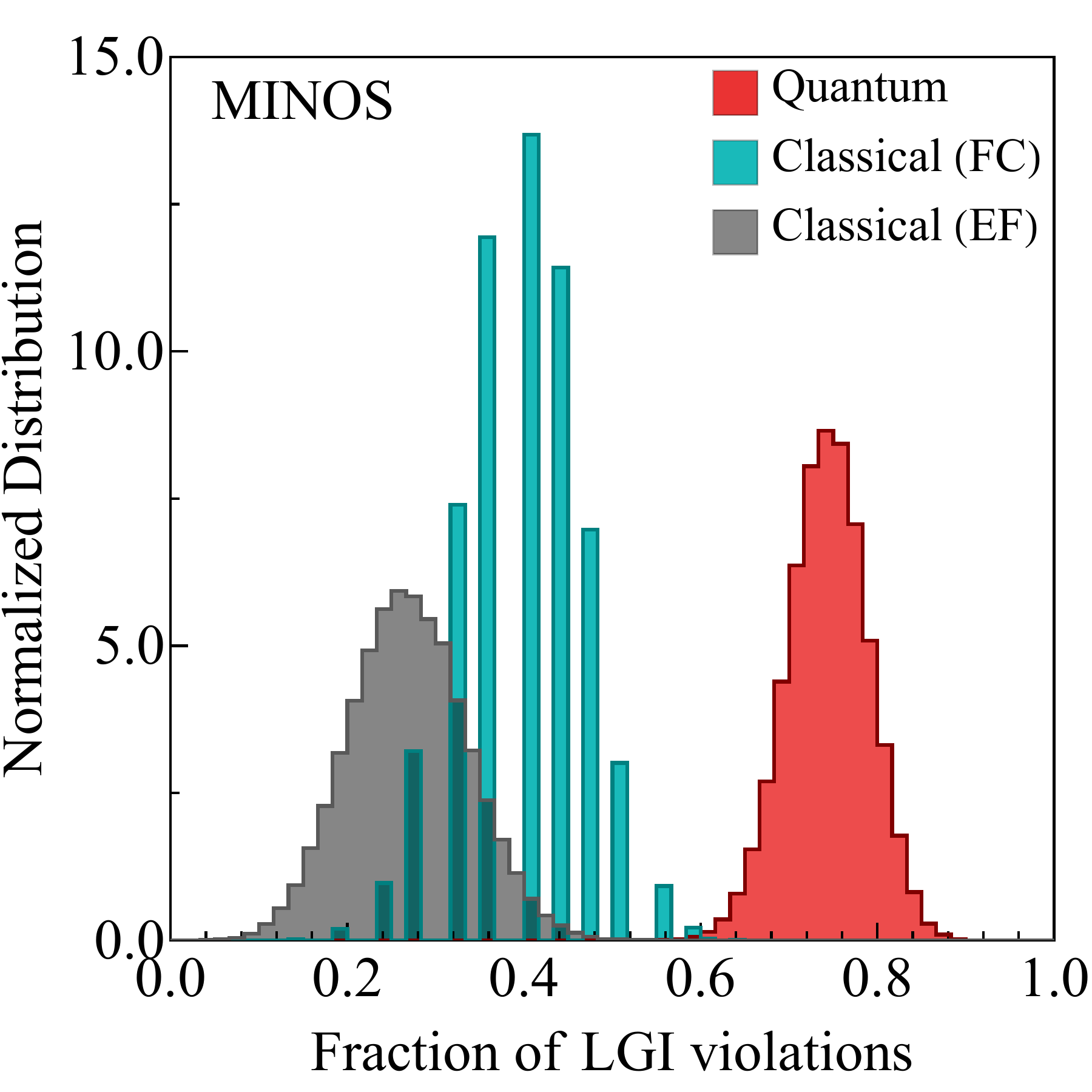}
    \end{minipage}
    \hfill
    \begin{minipage}{0.48\textwidth}
        \centering
           \includegraphics[width=\linewidth]{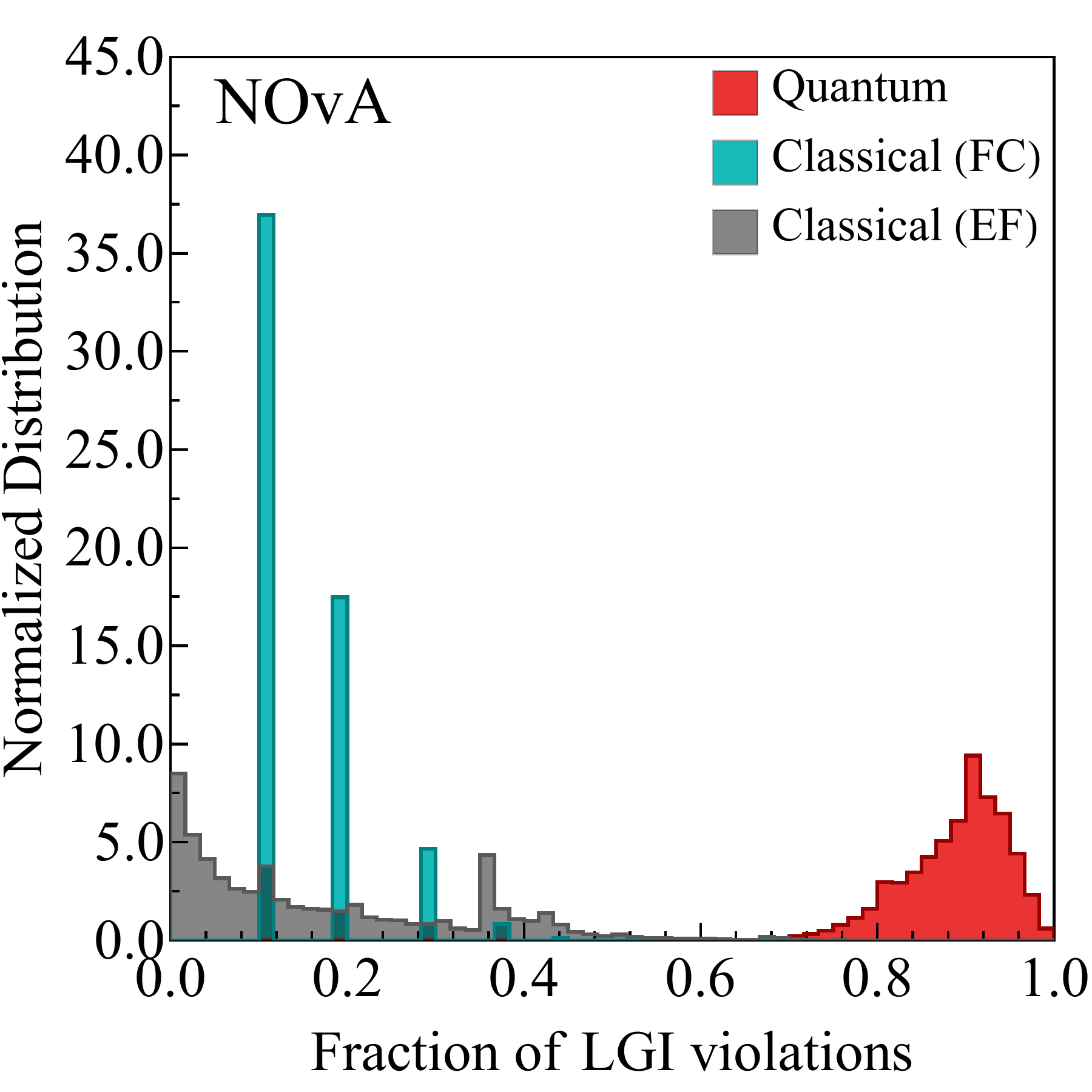}
    \end{minipage}
        \hspace{0.1cm}
        \centering
            \\ \vspace{0.2 cm}
    \begin{minipage}{0.48\textwidth}
        \centering
        \includegraphics[width=\linewidth]{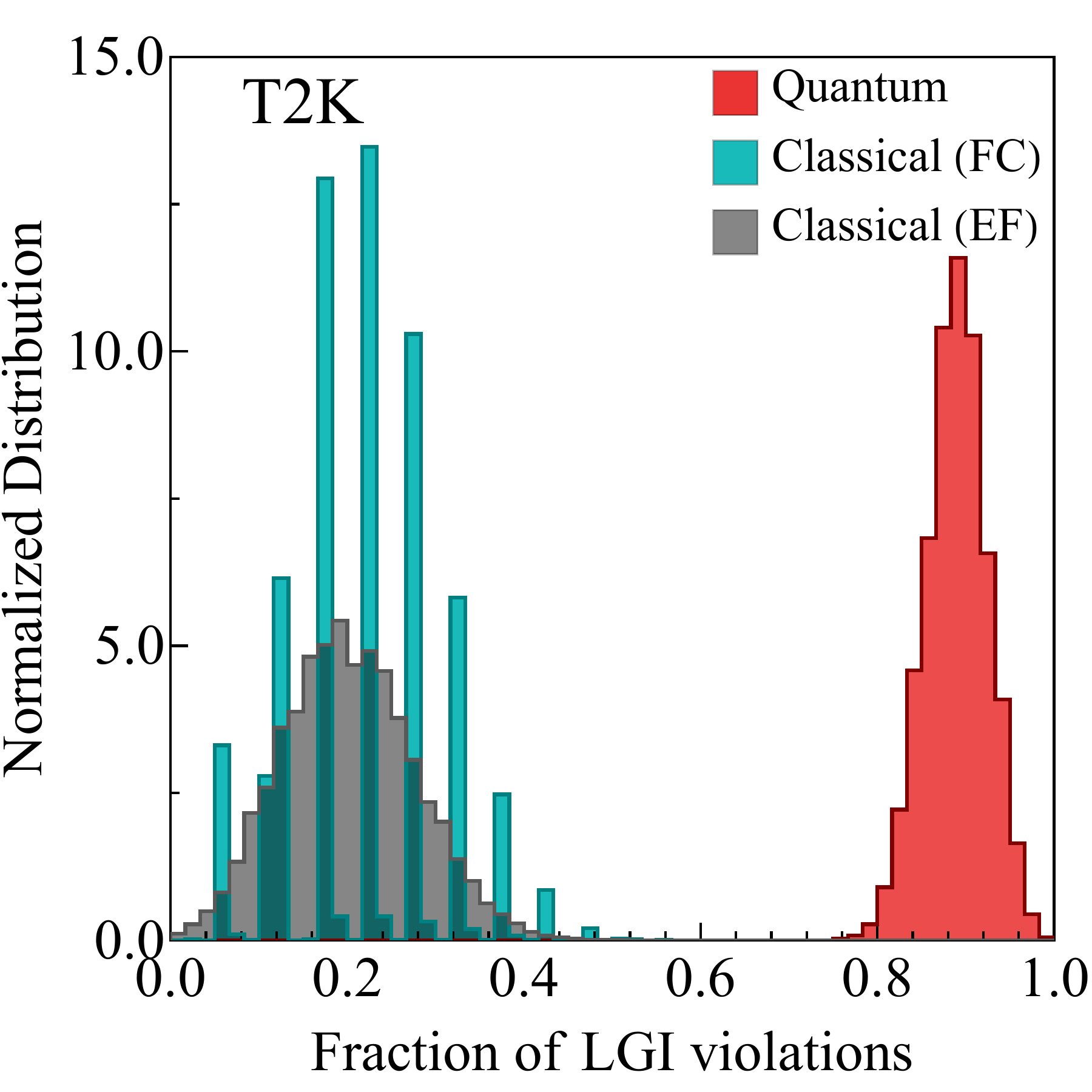}
    \end{minipage}
    \hfill
    \begin{minipage}{0.48\textwidth}
        \centering
        \includegraphics[width=\linewidth]{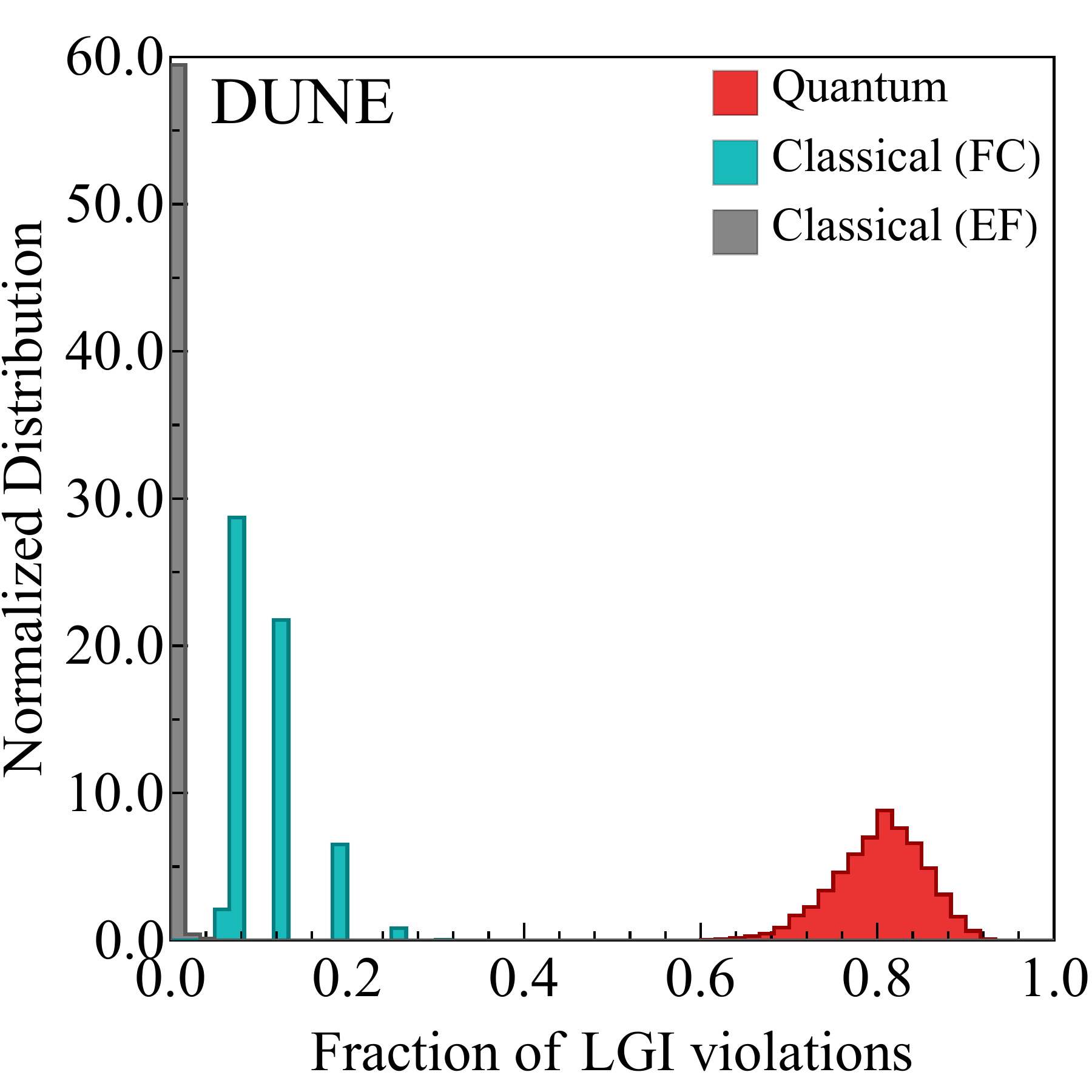}
    \end{minipage}
\caption{
Fraction of energy triplets that violate the Leggett-Garg inequality (LGI) for the four considered long-baseline experiments.
The red histograms correspond to the quantum scenario and show that a large fraction of triplets exceeds the classical bound.
The cyan histograms represent the classical factorized correlator (FC) method, employed for all 4 LG strings.
The gray histograms correspond to the classical exponential fit (EF) method.
Comparing the three cases, the quantum distributions clearly exhibit LGI violations, while the classical cases remain concentrated at comparatively smaller values.
}
    \label{fig:LGI_fraction_hist}
\end{figure}

\subsection{Fraction of Triplets that Violate the Leggett-Garg Inequality}

In this section, we compute the fraction of energy triplets that violate the LGI 
for each of the four considered experiments. The fraction of triplets violating LGI is simply calculated as the ratio of $N_{\rm LGV}$ and $N$, where $N$ is the total number of triplets.
For the quantum scenario, we evaluate the maximum of the four LG strings, $K_3^{\max}$, as defined in~\cref{eq:k3max}. We then compare these results against two classical cases, obtained through the FC and EF methods, which quantify the apparent LGI violations arising from statistical and systematic uncertainties.

For each energy bin, we sample the survival probability (or, in the EF method, the corresponding best-fit correlation function), including the effects of the experimental uncertainties, which are often asymmetric: see~\cref{sec:sampling} for a detailed discussion on sampling such asymmetric distributions using the split-Gaussian method.
We then form all possible energy triplets and compute the value of $K_3^{\max}$ for the quantum as well as for both classical scenarios (FC and EF). The fraction of triplets satisfying $K_3^{\max} > 1$ is calculated for each scenario.

In the red (quantum), dark cyan (classical FC), and gray (classical EF) histograms shown in~\cref{fig:LGI_fraction_hist}, we present the resulting fraction of LGI violations for all four experiments considered.
We sample the survival probabilities associated with each energy bin, and for every valid energy triplet combination, we compute the corresponding value of $K_3^{\max}$. This procedure is repeated $10^{5}$ times, yielding the normalized distributions shown in the figure.
All four experiments exhibit a very clear deviation from classical expectations, demonstrating the non-classical nature of neutrino oscillations.

Note that, in~\cref{fig:LGI_fraction_hist}, the histograms representing quantum behavior are approximately Gaussian for MINOS\footnote{Note that, for MINOS, our results do not fully match those from~\cite{Groth:2025gtf}, since we use the MINOS data from Ref.~\cite{Sousa:2015bxa, Holin:2015lya}, while the authors of~\cite{Groth:2025gtf} perform the sampling from the best-fit oscillation model for normal neutrino mass ordering, with the variance of this distribution approximated by the error bars of the MINOS data.}, T2K, and DUNE, but skewed to the right for NOvA.
For both the classical EF and FC distributions, only MINOS and T2K show Gaussian-like shapes, while NOvA and DUNE data lead to irregular non-Gaussian distributions. For the classical FC method, the sparse nature of the distribution leads to visible gaps in the histograms.

These features indicate that although the fraction of triplets violating the LGI, $f_{\rm LGV} \equiv N_{\rm LGV}/N$, gives a clear qualitative proof that neutrino oscillations are non-classical, this ratio is not an ideal quantitative measure for our purposes.
In particular, even though the distributions corresponding to the quantum scenario are clearly separated from the classical ones, the irregular and sparse character of the classical distributions makes it difficult to quantify the separation between the histograms corresponding to the classical and quantum scenarios.
Further, this measure does not incorporate the degree of LGI violation, as all violations are weighted equally.

Therefore, in what follows, we employ the $z$-score analysis, which is regulated by both: $(i)$ whether an energy triplet violates the LGI, and $(ii)$ the extent to which it exceeds the classical bound of $K_3 \leq 1$.
As we will observe, the distributions of the RMS $z$-score values provide a robust and quantitative measure of deviation from classicality.

\subsection{RMS \textit{z}-score}

\begin{figure}[t!]
    \centering
    \begin{minipage}{0.48\textwidth}
        \centering
           \includegraphics[width=\linewidth]{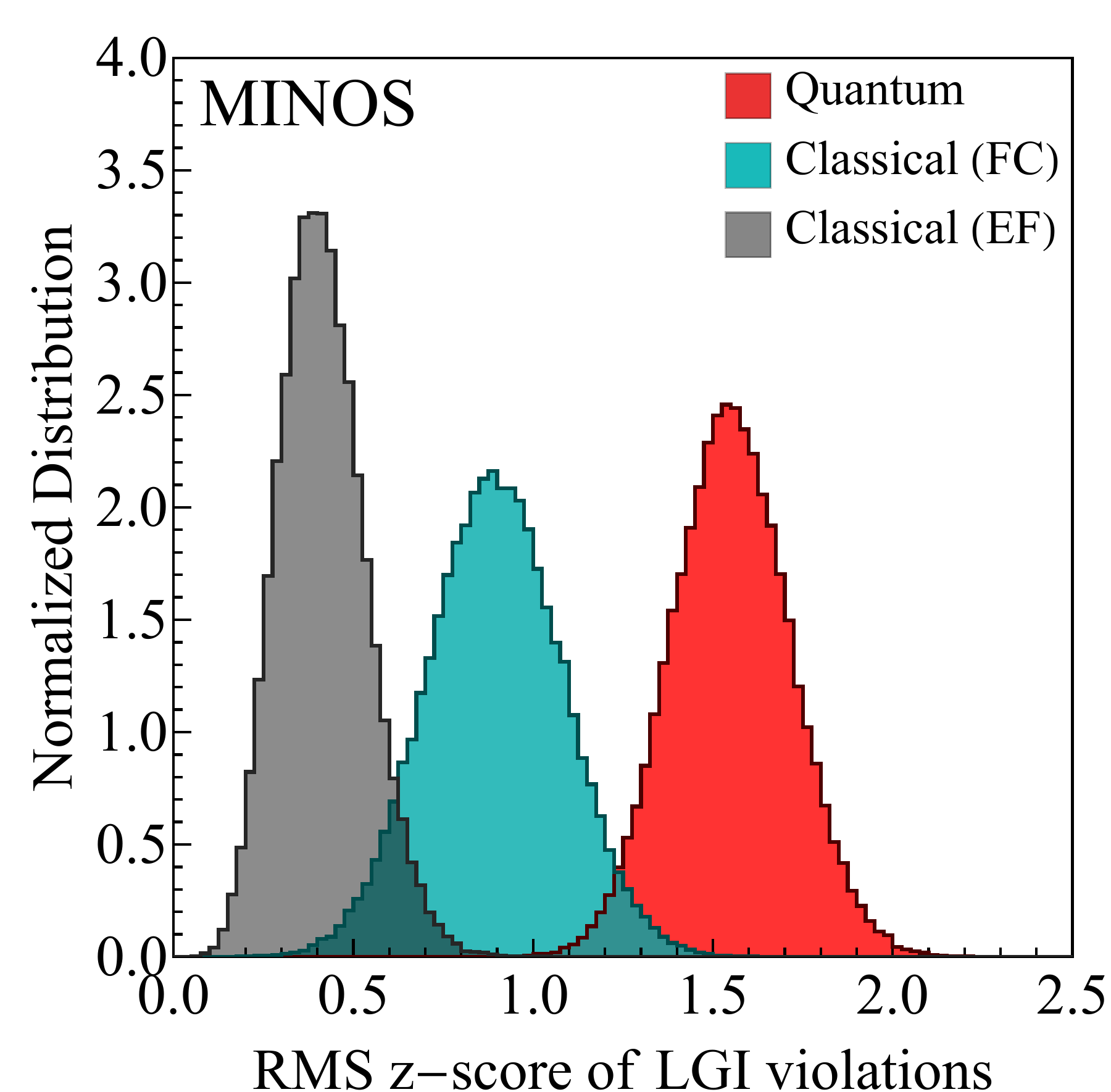}
    \end{minipage}
    \hfill
    \begin{minipage}{0.48\textwidth}
    \vspace{0.1cm}
        \centering
           \includegraphics[width=\linewidth]{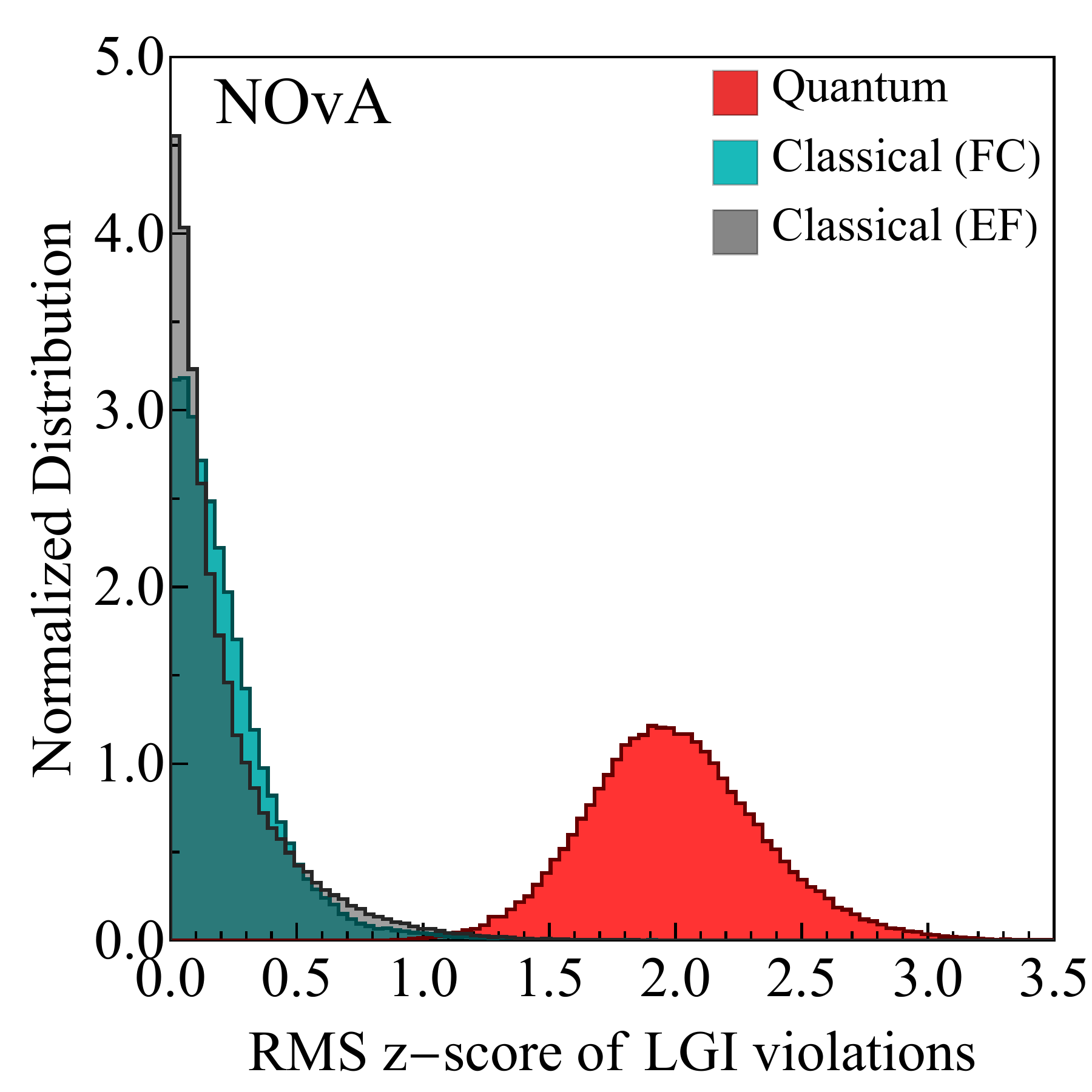}
    \end{minipage}
    \hspace{0.1cm}
    \\ \vspace{0.2 cm}
        \centering
    \begin{minipage}{0.48\textwidth}
        \vspace{0.1cm}
        \centering
        \includegraphics[width=\linewidth]{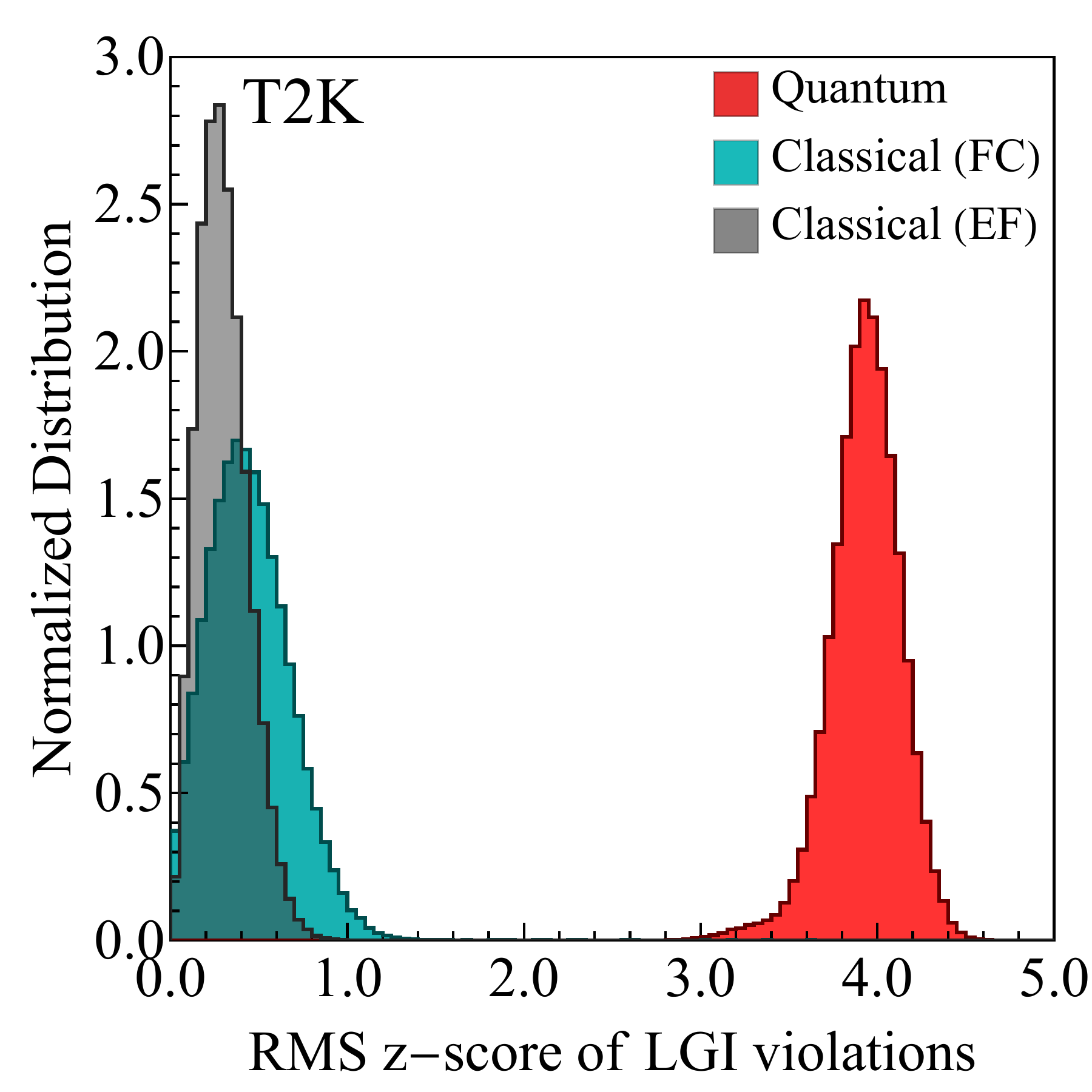}
    \end{minipage}
    \hfill
    \begin{minipage}{0.48\textwidth}
        \centering
        \includegraphics[width=\linewidth]{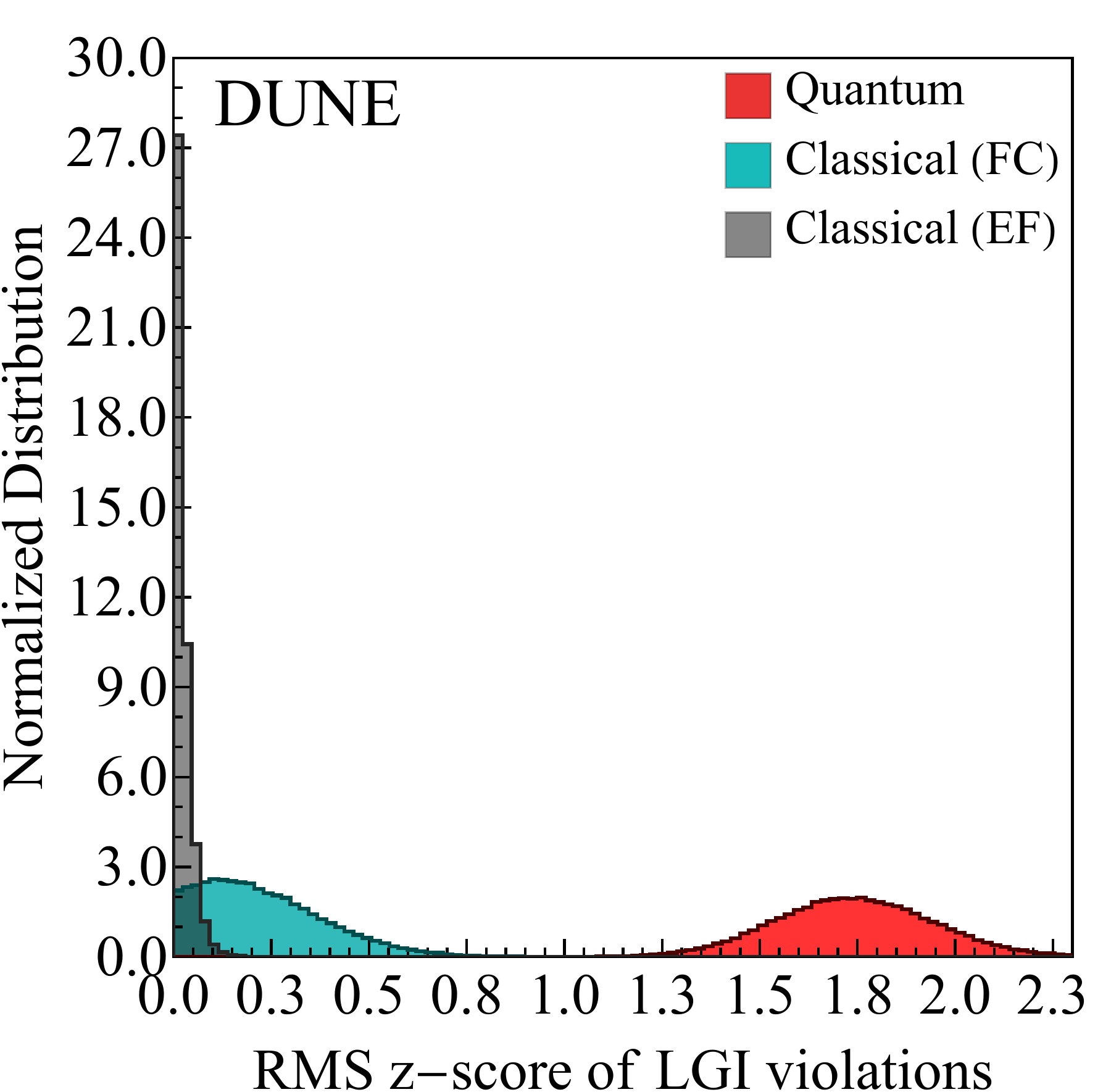}
    \end{minipage}
    \caption{
        Distributions of the RMS $z$-score $(z_{\rm RMS})$ for the four long-baseline experiments: MINOS, NOvA, T2K, and DUNE. The distributions are shown for the quantum (red), classical FC (cyan), and classical EF (gray) scenarios. For NOvA, T2K, and DUNE, the distributions corresponding to quantum scenarios are clearly separated from the classical ones, indicating a significant non-classical behavior.
    }
    \label{fig:zscore_all_expts}
\end{figure}

To obtain a more quantitative picture of the LGI violations in neutrino experiments, we employ the RMS $z$-score ($z_{\rm RMS}$) analysis, described in~\cref{sec:zscore}, see in particular~\cref{eq:zrms}.
In~\cref{fig:zscore_all_expts}, we present the $z_{\rm RMS}$ distributions for the four long-baseline experiments: MINOS, NOvA, T2K, and DUNE. Each panel shows the results for the quantum scenario (red) alongside the two classical baselines: FC (cyan) and EF (gray). 

In the top left panel of~\cref{fig:zscore_all_expts}, we show the $z_{\rm RMS}$ distributions for the MINOS experiment. Note that, unlike NOvA and T2K, the choice of classical modeling has a strong impact on MINOS: the EF (gray) and FC (cyan) distributions are very clearly separated from each other.
We also observe that, among the four experiments, MINOS exhibits the smallest separation between the quantum and classical scenarios (comparing the cyan/gray and red histograms).
In~\cref{fig:zscore_all_expts}, our results for MINOS differ from Ref.~\cite{Groth:2025gtf}. This is because:
\begin{enumerate}
    \item We employ a modified definition of $z_{\rm RMS}$, normalizing against the total number of energy triplets ($N$), instead of the number of triplets that violate the LGI ($N_{\rm LGV}$), see discussions in~\cref{sec:z_rms};
    \item As discussed previously, the authors of Ref.~\cite{Groth:2025gtf} sample the data with the mean defined from the best fit, and the variance taken from the MINOS data, while we sample both the mean and the variance from the data in~\cite{Sousa:2015bxa, Holin:2015lya};
    \item We remind the reader that we also employ a new method for constructing the viable energy triplets, capturing the realistic experimental energy bin widths, see~\cref{sec:energy_triplets}.
\end{enumerate}
We are able to perfectly reproduce the results from Ref.~\cite{Groth:2025gtf} in the absence of the modifications discussed above.

In the top right panel of~\cref{fig:zscore_all_expts}, we show the results for NOvA,  where the distance between the classical and quantum distributions is also comparatively small relative to the bottom two panels (T2K and DUNE).

As can be observed in the bottom panels of~\cref{fig:zscore_all_expts}, the $z_{\rm RMS}$ measure is particularly effective at highlighting the non-classical nature of neutrino oscillations in T2K and DUNE.
For both experiments, the separation between the quantum and classical distributions is very pronounced.

Further, note that for NOvA, T2K, and DUNE, the classical distributions deviate from a Gaussian shape, primarily due to the RMS nature of the measure, which does not allow the $z_{\rm RMS}$ measure to take a value smaller than zero.
Note, however, that the strong sparse nature seen earlier in the $f_{\rm LGV}$ classical FC distributions in~\cref{fig:LGI_fraction_hist} is no longer present in the $z_{\rm RMS}$ analysis. This allows us to fit Gaussian PDFs or CDFs to the resulting quantum and classical distributions and extract best-fit mean values and standard deviations for each scenario.

In what follows, we discuss in detail how the $z_{\rm RMS}$ distributions shown in~\cref{fig:zscore_all_expts} can be quantified, including the key steps involved in fitting Gaussian PDFs or CDFs, the criteria for choosing between them, and the possible limitations involved.

\subsection{Quantifying the Distributions}

As shown in~\cref{fig:zscore_all_expts}, all the $z_{\rm RMS}$ distributions for the quantum scenario are approximately Gaussian and can be well described by a standard probability distribution function (PDF) fit.
However, the classical baselines (obtained with both the FC and EF method), particularly for NOvA, T2K, and DUNE, remain close to zero and exhibit non-Gaussian shapes.
In this section, we describe how to treat the non-Gaussian cases using an effective CDF-based fitting method.

\subsubsection{Fitting the Data with Gaussian PDFs}

Among the classical EF and FC distributions for the considered long-baseline experiments, only the results for MINOS can be reliably fit with a Gaussian PDF to extract robust best-fit mean values and standard deviations. In contrast, both NOvA and DUNE classical distributions are distinctly non-Gaussian, while the T2K classical distribution is of a partial Gaussian shape.

Nevertheless, even though a fully Gaussian shape is absent, we can still attempt a simple Gaussian PDF fit. Applying such a fit to the NOvA results, we obtain the following best-fit parameters for the two classical cases and for the quantum distribution:
\begin{align}
    \mu_{\rm FC} = 0.223 \,,\; \sigma_{\rm FC} = 0.203 \, , & \qquad \mu_{\rm EF} = 0.221 \, ,\; \sigma_{\rm EF} = 0.239 \, , \nonumber\\
    \mu_{\rm Q} = 1.996 \, ,& \;  \sigma_{\rm Q} = 0.348  \; ,
    \label{eq:nova_pdf_fit}
\end{align}
with $\rm{Q}$ denoting the quantum distribution.
However, such a PDF-based fit suffers from two important limitations:
\begin{enumerate}
    \item Robustness of the fit: For non-Gaussian or weakly peaked distributions, the Gaussian fit may not accurately represent the underlying shape. The resulting best-fit parameters can be highly sensitive to fluctuations in the right tail or to minor changes near the peak, leading to unstable or misleading estimates of the mean.
    \item Physical interpretation: A naive Gaussian PDF fit may give a negative best-fit mean for the $z_{\rm RMS}$ distribution; this is unphysical given that $z_{\rm RMS}$ is strictly non-negative by construction, see~\cref{eq:zrms}.
\end{enumerate}

Therefore, we devise an effective CDF-based fitting method, motivated by the observed distribution and more appropriate for our primary aim: quantifying how far the quantum distribution lies from the classical baselines.

\subsubsection{Fitting the Data with Gaussian CDFs}

\vspace{2em}
\begin{figure}[t!]
    \centering
    \begin{minipage}{0.45\textwidth}
        \centering
        \includegraphics[width=\linewidth]{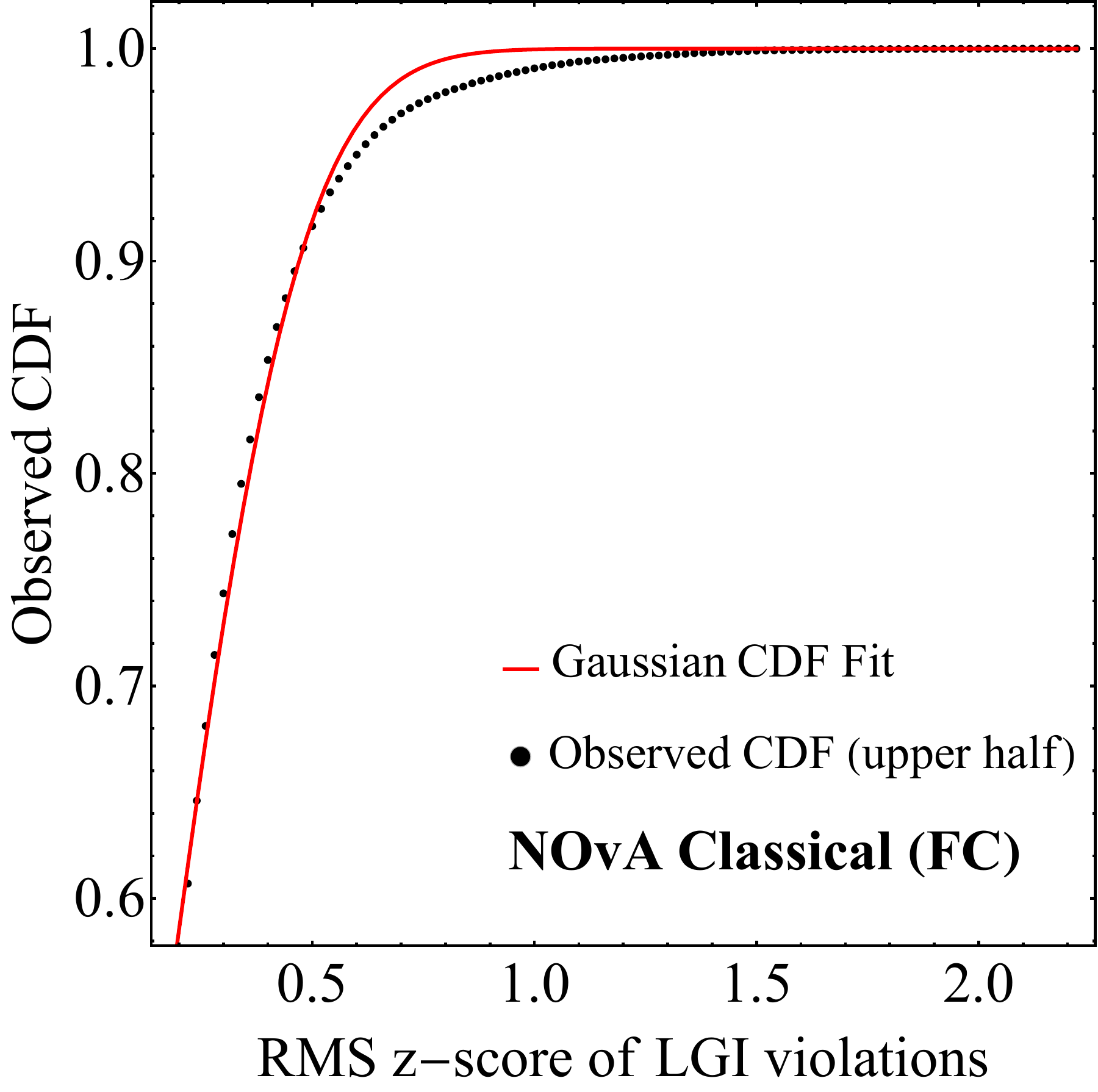}
    \end{minipage}
    \hfill
    \begin{minipage}{0.45\textwidth}
        \centering
        \includegraphics[width=\linewidth]{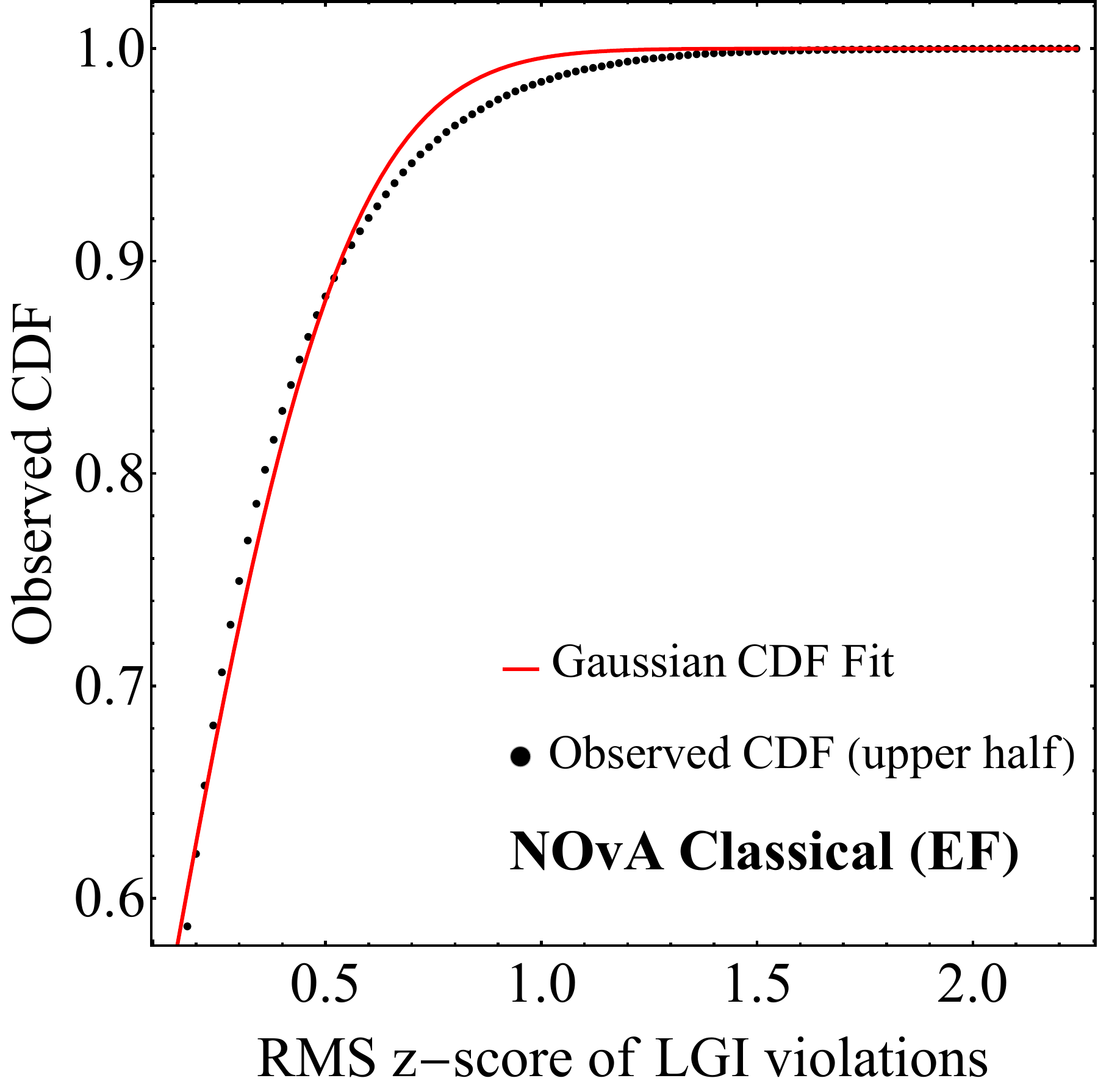}
    \end{minipage}
    \\
    \vspace{0.8cm}
    \begin{minipage}{0.45\textwidth}
        \centering
        \includegraphics[width=\linewidth]{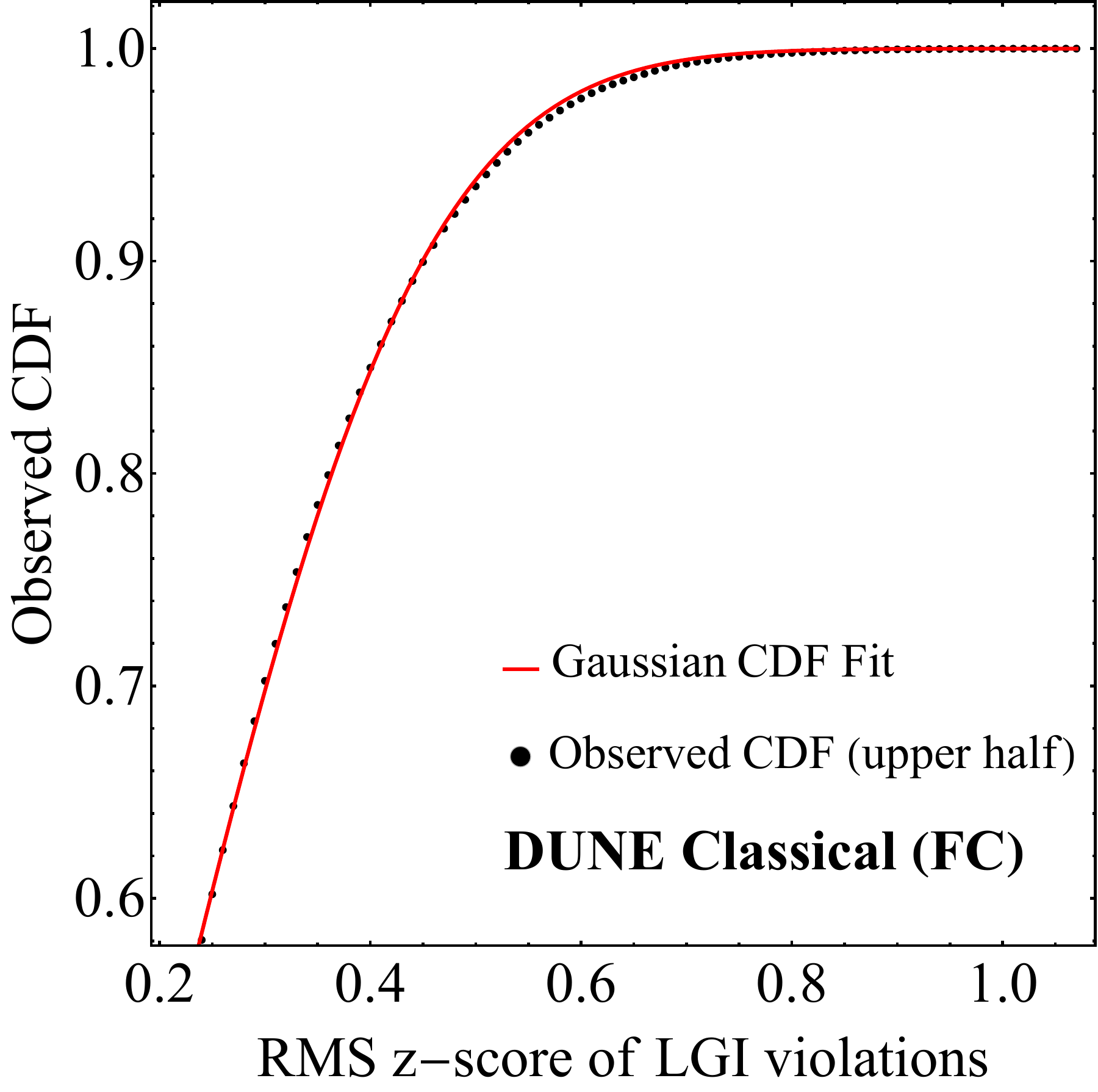}
    \end{minipage}
    \hfill
    \begin{minipage}{0.45\textwidth}
        \centering
        \includegraphics[width=\linewidth]{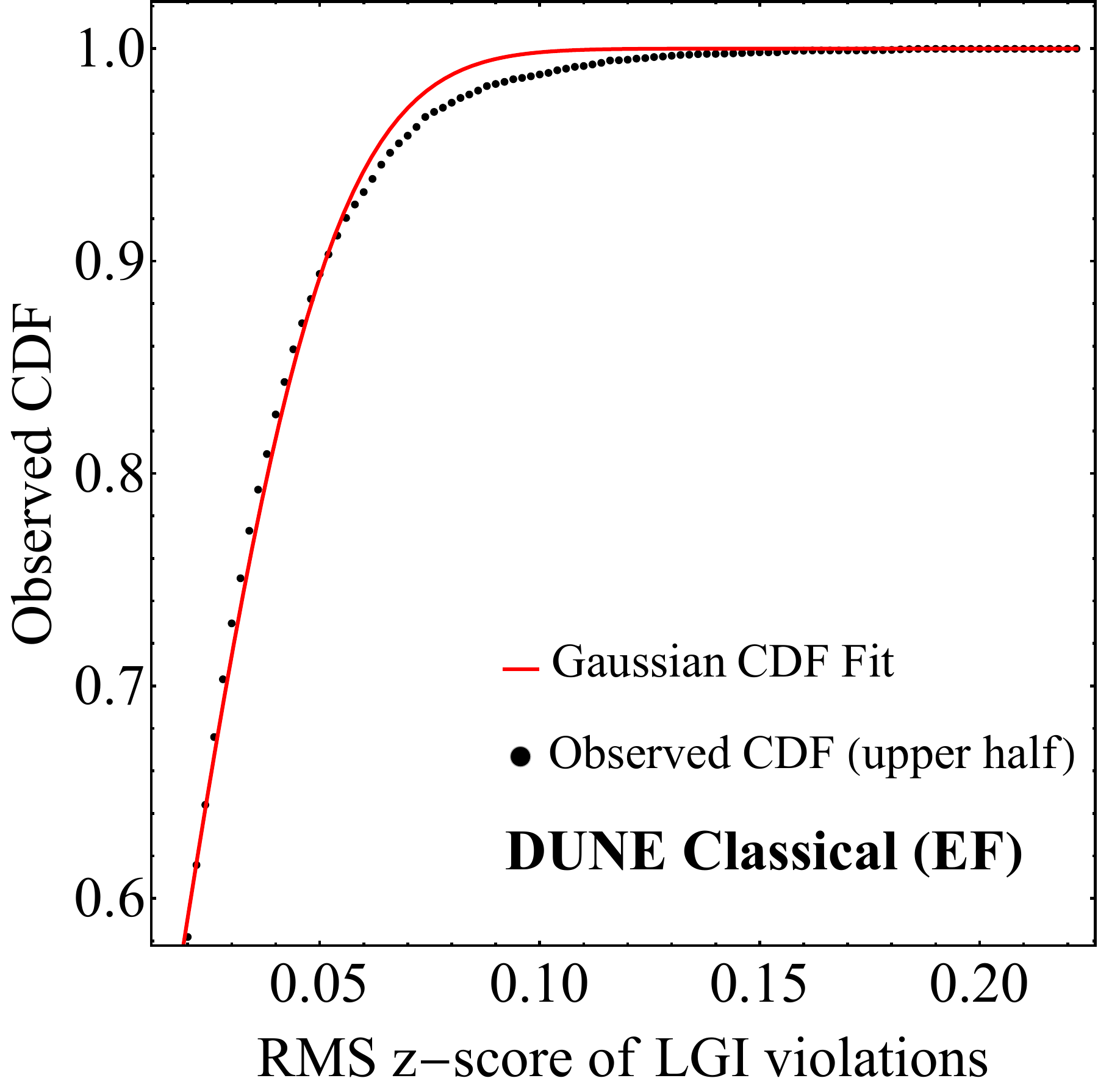}
    \end{minipage}
    \caption{
        Empirical CDFs (black dotted) for NOvA (top panels) and DUNE (bottom panels) classical $z_{\rm RMS}$ scores, with a best fit (red line) obtained from the upper half of the data using a Gaussian CDF fit method.
    }
    \label{fig:nova_dune_cdfs}
\end{figure}

To fit the classical distributions using an effective CDF-based approach, we proceed as follows:
\begin{enumerate}
    \item Using distributions in~\cref{fig:zscore_all_expts}, we compute the empirical CDFs for the classical $z_{\rm RMS}$ values. The empirical CDFs, therefore, give the fraction of samples with $z$-scores less than or equal to a given value.

    \item We focus on the upper half of the data (i.e., $z$-scores above the median, corresponding to CDF values $\geq 0.5$) and fit this portion with a Gaussian CDF, allowing both the mean ($\mu$) and standard deviation ($\sigma$) to vary. We observe that the Gaussian CDF best fit yields a smooth and stable representation of the distribution, even when the full shape is non-Gaussian and asymmetric.

    \item The resulting best-fit parameters $(\mu, \sigma)$ are extracted separately for the two classical distributions (FC and EF), for each long-baseline experiment.
\end{enumerate}
This method is particularly robust because $(i)$ it relies only on the upper 50\% of the data, where the distributions behave more smoothly, and $(ii)$ it captures the physically relevant right tail, whose proximity to the quantum distribution is important for quantifying how closely the classical scenarios can mimic the quantum behavior.

In~\cref{fig:nova_dune_cdfs}, we show the results of this procedure for the classical (EF and FC) distributions in NOvA (upper panels) and DUNE (lower panels). The black points indicate the empirical CDFs of the classical samples, while the red curves represent the best-fit Gaussian CDFs obtained from fitting the upper 50\% of the data. 
The close agreement between the data and the best fit demonstrates the effectiveness of this method, even when the corresponding histograms are non-Gaussian.
For NOvA, the Gaussian CDF fits yield:
\begin{align}
    \mu_{\rm FC} = 0.146 \,,\; \sigma_{\rm FC} = 0.254 \, ; & \qquad \mu_{\rm EF} = 0.088 \, ,\; \sigma_{\rm EF} = 0.348 \, .
    \label{eq:nova_cdf_fit}
\end{align}
These values are in moderate agreement with those obtained from the simple PDF fit (see~\cref{eq:nova_pdf_fit}); Note that both the $\sigma_{\rm FC}$, and $\sigma_{\rm EF}$, obtained through the effective CDF fit, are broader than the values obtained through the PDF fit.

In~\cref{tab:fit_values}, we show the best-fit Gaussian parameters (mean and standard deviation) obtained for all 3 histograms (2 classical + 1 quantum) and for the 4 considered experiments.
We highlight the DUNE and NOvA classical EF and FC fits with cyan, where the best fits are obtained through the CDF method. The rest are highlighted with light brown, where the Gaussian PDF fit method is employed.
With robust best-fit parameters obtained for all quantum and classical distributions, one can quantify how strongly each experiment departs from classicality, i.e., how quantum the behavior of neutrino oscillations is in each experiment.

\begin{table}[t]
    \centering
    \begin{tabular}{||c||c|c||c|c||c|c||}
    \hline
    \hline
        Experiment & $\mu_Q$  & $\sigma_Q$ & $\mu_{\rm FC}$ & $\sigma_{\rm FC}$ & $\mu_{\rm EF}$ & $\sigma_{\rm EF}$\\
    \hline
    \hline
    MINOS & \cellcolor[HTML]{F3D8C7} 1.551  & \cellcolor[HTML]{F3D8C7} 0.164 & \cellcolor[HTML]{F3D8C7} 0.894 & \cellcolor[HTML]{F3D8C7} 0.185 & \cellcolor[HTML]{F3D8C7} 0.414 & \cellcolor[HTML]{F3D8C7} 0.120  \\
    \hline
    NOvA & \cellcolor[HTML]{F3D8C7} 1.996  & \cellcolor[HTML]{F3D8C7} 0.348 & \cellcolor[HTML]{BEE1E6} 0.146 & \cellcolor[HTML]{BEE1E6} 0.254 &\cellcolor[HTML]{BEE1E6} 0.088 & \cellcolor[HTML]{BEE1E6} 0.348  \\
    \hline
    T2K & \cellcolor[HTML]{F3D8C7} 3.929  & \cellcolor[HTML]{F3D8C7} 0.204 & \cellcolor[HTML]{F3D8C7} 0.450 & \cellcolor[HTML]{F3D8C7} 0.251 & \cellcolor[HTML]{F3D8C7} 0.298 & \cellcolor[HTML]{F3D8C7} 0.139   \\
    \hline
    DUNE$^\dagger$ & \cellcolor[HTML]{F3D8C7} 1.742  & \cellcolor[HTML]{F3D8C7} 0.202 & \cellcolor[HTML]{BEE1E6} 0.198 & \cellcolor[HTML]{BEE1E6} 0.194 & \cellcolor[HTML]{BEE1E6} 0.013 & \cellcolor[HTML]{BEE1E6} 0.029  \\
    \hline
    \hline
    \end{tabular}\\
    \small{$^\dagger$Uncertainty consisting of statistical error$\,$+$\,$an uniform 5\% systematic error for each energy bin.}
    \caption{The mean ($\mu$) and standard deviation ($\sigma$) for each of the distributions corresponding to classical (FC and EF) and quantum scenarios, for the four considered long-baseline experiments, obtained through the Gaussian PDF (light brown) and effective CDF (cyan) fit methods.}
    \label{tab:fit_values}
\end{table}

\subsection{Comparing the Significance of Quantumness Across Experiments}

We compute the significance of deviations from classical realism for each experiment by using the following formula:
\begin{equation}
    N_{\rm Q-Classical}^{(\sigma)} = \frac{|\mu_{\rm Q} - \mu_{\rm Classical}|}{\sigma_{\rm Classical}} \; .
    \label{eq:significance}
\end{equation}
Here, $\mu_{\rm Q}$ is the best-fit mean of the quantum distribution, when fitted with a Gaussian distribution, $\mu_{\rm Classical}$ and $\sigma_{\rm Classical}$ are the mean and standard deviation of the Gaussian best fit corresponding to a given classical distribution.

In~\cref{tab:significance}, we show the statistical significance of quantum behavior for each of the four long-baseline neutrino experiments: MINOS, NOvA, T2K, and DUNE.
The final significance for quantumness is impacted by several factors: the measured oscillation probabilities and their uncertainties, the total range of neutrino energies, the total number of energy triplets, the number of data points, as well as the choice of classical baselines for comparison against the quantum scenario.
The more conservative result, between the separation of $(i)$ quantum and classical FC predictions, and $(ii)$ quantum and classical EF predictions, is highlighted in blue.
Note that comparing against the classical FC method leads to a more conservative statistical significance for MINOS and T2K data, and DUNE benchmark. In contrast, comparing against the classical EF method gives a more conservative answer for NOvA.

\begin{table}[t]
    \centering
    \begin{tabular}{||c||c|c||}
    \hline
    \hline
       \quad Experiment \; \qquad & \quad $N_{\rm Q-FC}^{(\sigma)}$ \; \qquad & \quad  $N_{\rm Q-EF}^{(\sigma)}$  \; \qquad \\
    \hline
    \hline
    MINOS & \cellcolor[HTML]{bcd4e6} 3.55$\sigma$ & 9.48$\sigma$ \\
    \hline
    NOvA & 7.28$\sigma$ & \cellcolor[HTML]{bcd4e6} 5.48$\sigma$ \\
    \hline
    T2K & \cellcolor[HTML]{bcd4e6} 13.86$\sigma$ & 26.12$\sigma$ \\
    \hline
    DUNE & \cellcolor[HTML]{bcd4e6} 7.98$\sigma$ & 58.90$\sigma$ \\
    \hline
    \hline
    \end{tabular}
    \caption{Statistical significance of quantumness for each of the four considered long-baseline neutrino experiments. The two columns present results for comparison of the quantum distribution against the classical EF and FC cases. The DUNE results correspond to a benchmark with 5\% systematic error. For each experiment, we highlight the more conservative result in blue.}
    \label{tab:significance}
\end{table}

\begin{figure}[t!]
\centering
\includegraphics[width=0.458\linewidth]{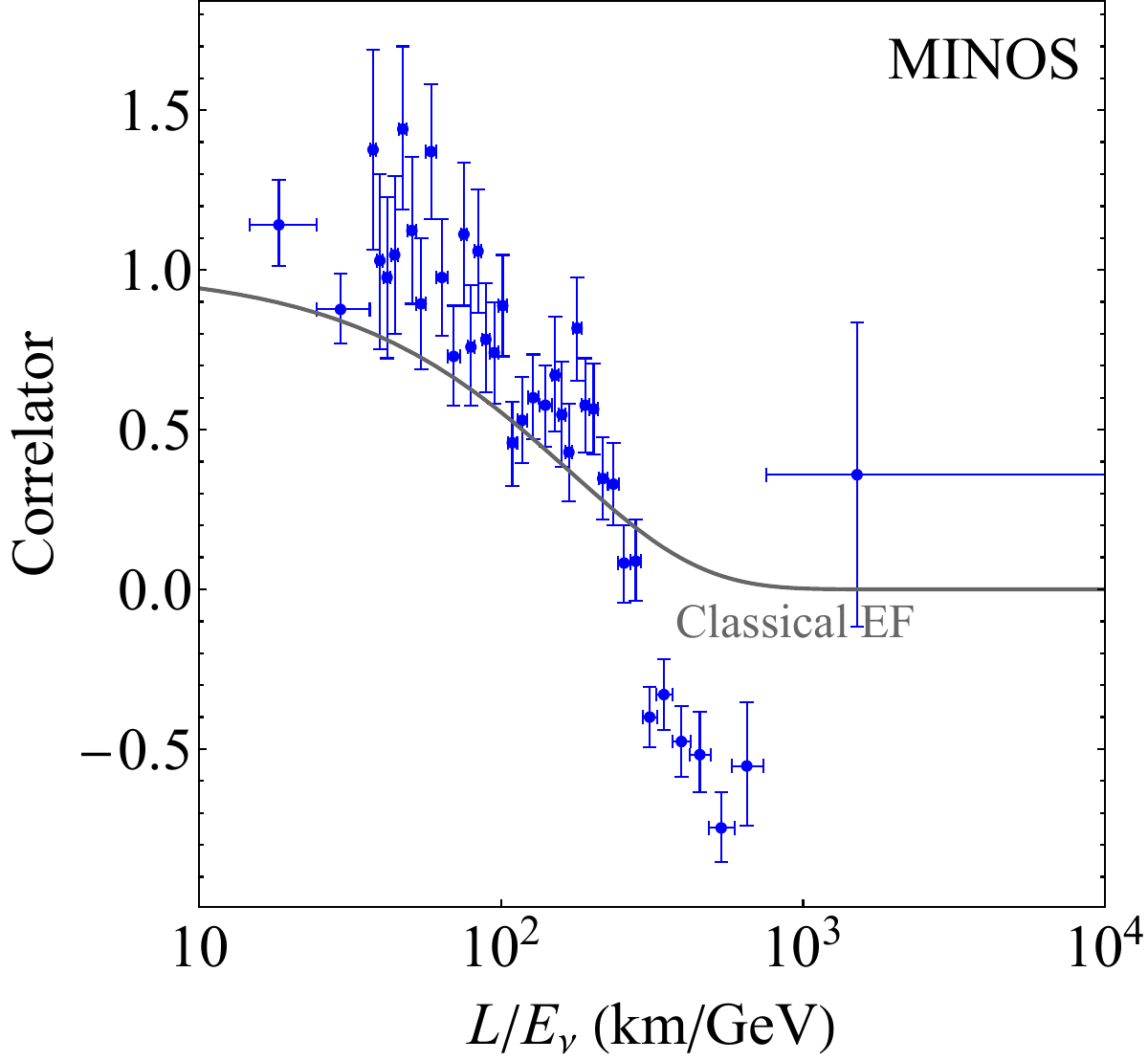}
\hfill
\includegraphics[width=0.44\linewidth]{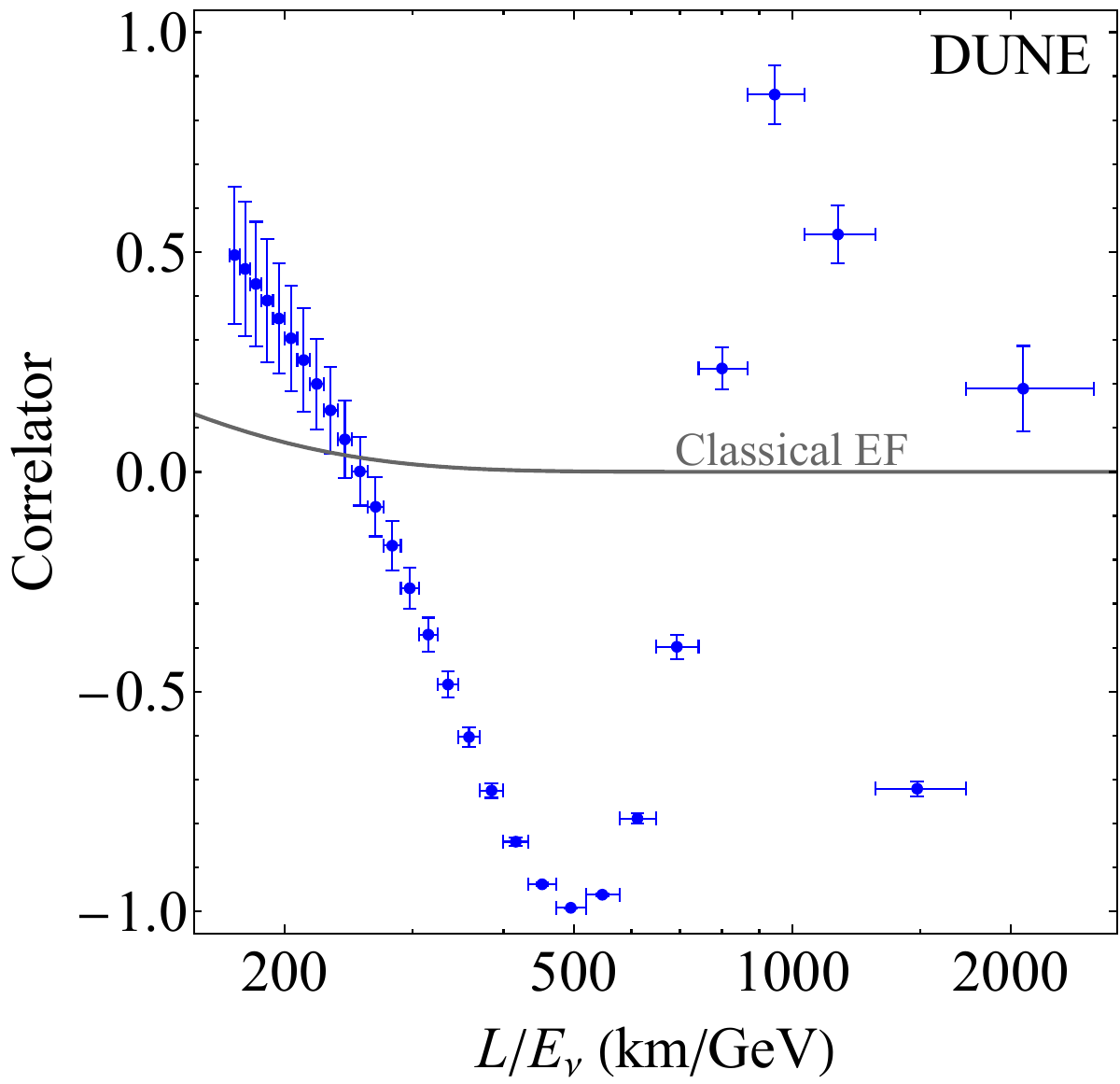}
\caption{Classical EF best fit for the correlators, for MINOS (left panel) and DUNE (right panel). The gray curve denotes the Classical EF $C_{ij}$, with the correlator values obtained from the experimental survival probability data shown in blue. For MINOS, the best-fit value is $\Gamma = 5.90\times 10^{-3}~\rm{GeV}/\rm{km}$, and for DUNE, the best-fit curve is obtained for $\Gamma = 1.35\times 10^{-2}~\rm{GeV}/\rm{km}$.}
\label{fig:classical_EF_best_fit}
\end{figure}

The comparatively larger values of $N_{\rm Q\text{-}EF}^{(\sigma)}$ in~\cref{tab:significance} arise because the classical EF $z_{\rm RMS}$ distribution has a smaller mean $\mu_{\rm EF}$ and a narrower width $\sigma_{\rm EF}$. 
Specifically, the large value of $N_{\rm Q\text{-}EF}^{(\sigma)} \approx 58.9\sigma$ for DUNE can be explained qualitatively through~\cref{fig:classical_EF_best_fit}, where we show the correlators ($C_{ij}$) for the best fit obtained via the classical EF method, for MINOS, and DUNE. 
The smallness of the best fit value $\Gamma = 5.90\times 10^{-3}~\rm{GeV}/\rm{km}$ for MINOS allows the classical EF correlator to have values close to 1 (see left panel of~\cref{fig:classical_EF_best_fit}), especially at lower $L/E_\nu$.
However, for DUNE we obtain $\Gamma = 1.35\times 10^{-2}~\mathrm{GeV}/\mathrm{km}$, which leads to a much smaller correlator. Moreover, the smaller uncertainty on the muon neutrino survival probability for DUNE directly translates into a smaller uncertainty on the correlator $C_{ij}$. For DUNE, unlike MINOS, the near-zero values of $C_{ij}$ with small uncertainties make the classical EF method less likely to violate the LGI, and even when this happens, the violation is typically small. As a result, the $z_{\rm RMS}$ distribution for DUNE has a small mean and standard deviation, which leads to a large value of $N_{\rm Q\text{-}EF}^{(\sigma)}$.

Among the three existing experiments, we observe the largest significance of quantum behavior at T2K. We would like to stress that following the analysis strategy outlined in~\cite{Formaggio:2016cuh, Groth:2025gtf}, we have obtained a comparably large significance for T2K:
\begin{align}
    N_{\rm Q-FC,\, tol.}^{(\sigma)} \approx 13.1\sigma \; , \qquad     N_{\rm Q-EF,\, tol.}^{(\sigma)} \approx 22.9\sigma\; .
    \label{eq:significance_tol}
\end{align}

Given that we found a comparable number of energy triples as in \cite{Groth:2025gtf} (for a phase tolerance of $5\%$), $\sim 90\%$ of which coincide across two analyses, the energy triplet selection method clearly does not govern the overall sensitivity. Instead, the large significance stems from the asymmetric uncertainties in the data points, which is characteristic of the analyzed T2K data \cite{patrick_dunne_2020_3959558}, but not of MINOS data \cite{Sousa:2015bxa, Holin:2015lya}.
For T2K, we infer this by considering pseudodata with symmetric uncertainties. Using the upper (lower) error bars in the oscillation probability data as a proxy typically corresponds to larger (smaller) uncertainties. We observe that in the presence of larger (smaller) uncertainties, the mean value of the  $z_{\rm RMS}$ quantum distribution is at $\mu_{Q} \approx 2$ ($\mu_{Q} \approx 4$). This is because the denominator in the $z$-score definition is directly affected by the width of the error bars (see~\cref{sec:z_score_defn}). In fact, for upper (lower) uncertainties in oscillation probability data, we note that the significance of quantumness becomes smaller (larger) than that in~\cref{eq:significance_tol}. Therefore, our split-Gaussian sampling method, which can address asymmetric uncertainties in oscillation probability data, is ideal for experiments like T2K.
We remind the reader that this work is the first one to analyze the T2K data in the context of quantumness.

\subsection{DUNE Results with Varying Systematic Error}
\label{sec:varying_systematic}

We now investigate how the significance of quantumness (defined as the separation between the distributions corresponding to quantum and classical $z_{\rm RMS}$ behaviors) for DUNE changes when a varying systematic uncertainty is included on top of the $1/\sqrt{N_{E_\nu}}$ statistical errors.

For each energy bin $(E_\nu)$, the total error is given by $ \delta P (E_\nu) = [ (\delta P_{\rm syst})^2 + (\delta P_{\rm stat})^2 ]^{1/2} $, where $\delta P_{\rm stat}$ is the statistical error obtained from~\cref{eq:stat_error}.
We remind the reader that in order to obtain the pseudodata, we use the parameters given in~\cref{eq:oscparam}, and follow the same procedure as previously outlined at the beginning of~\cref{sec:results}.
In our treatment, we take the contribution of the systematic error to be the same for each energy bin.

\begin{figure}[t!]
\centering
\includegraphics[width=0.5\linewidth]{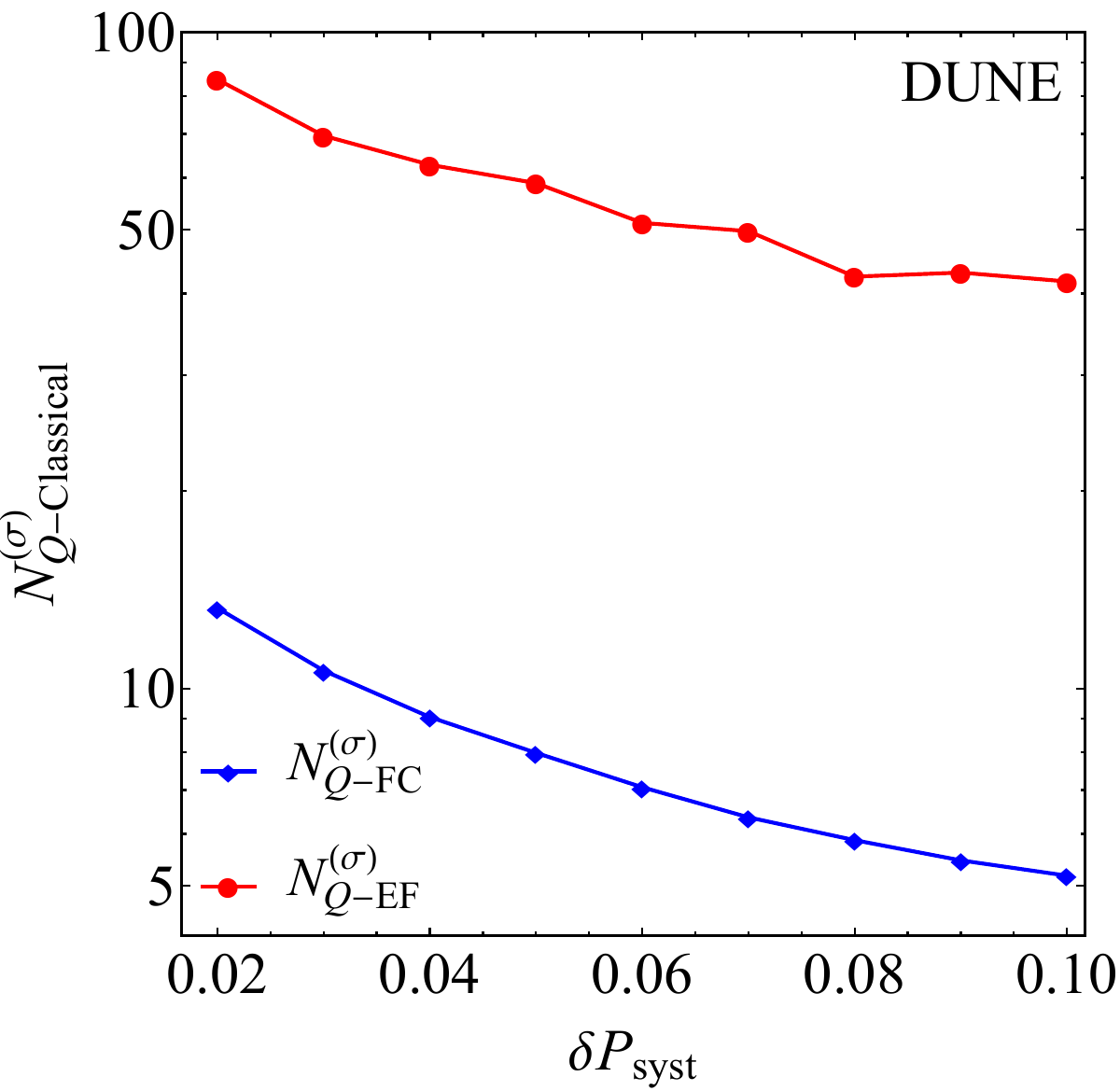}
\caption{Significance of quantumness at DUNE, compared against the two classical baselines (FC and EF), as a function of the systematic error $(\delta P_{\rm syst})$.}
\label{fig:dune-sigma}
\end{figure}

We vary the systematic error $(\delta P_{\rm syst})$ in the range $(0.02,0.1)$.
From the resulting quantum and classical $z_{\rm RMS}$ distributions, we extract their mean values and standard deviations and calculate their separation using~\cref{eq:significance}. The results are shown in~\cref{fig:dune-sigma}.
The separation between the quantum and classical scenarios gradually decreases as the systematic error grows.
However, even at the 10\% level ($\delta P_{\rm syst} = 0.1$), the quantum prediction remains several sigma away from both classical constructions, with the classical FC baseline consistently producing more conservative results across the whole range of $\delta P_{\rm syst}$ values. The significance of the classical EF baseline is large due to the extremely narrow classical distribution for the EF scenario.
Note that, even for a systematic uncertainty of $\delta P_{\rm syst} \approx 2\%$, the DUNE significance of quantumness cannot surpass the T2K result.

\subsection{Summary of the Results}

In summary, the $z_{\rm RMS}$ measure employed makes the separation between the distributions corresponding to quantum and classical scenarios easier to observe and quantify. MINOS, NOvA, T2K, and DUNE all show a significant statistical deviation between the quantum and classical distributions, with T2K giving the strongest result at
\begin{equation}
    N_{\rm Q-FC}^{(\sigma)}\Big|_{\rm T2K} = 13.86\sigma \; ,
\end{equation}
for the more conservative scenario (quantum vs. classical FC). The main numerical results can be found in~\cref{tab:fit_values,tab:significance}.

For DUNE, by varying the assumed systematic uncertainty, we find that the significance of quantumness increases with decreasing systematic errors. This behavior is expected and may play an important role in determining the eventual significance of quantum effects observed at DUNE.
Overall, our results show that all four long–baseline neutrino experiments exhibit a significant violation of classical realism, and some experiments, like T2K, are clearly more sensitive to the violations of the Leggett–Garg inequality.

\section{Conclusions}
\label{sec:conclusion}

Neutrino experiments provide an opportunity to study the quantum behavior of nature over hundreds of kilometers of length scales.
In this work, we have shown that violations of classicality can be tested in long-baseline neutrino oscillation experiments. We considered four experiments: MINOS, NOvA, T2K, and DUNE. For each of them, we studied the violations of the Leggett-Garg inequality (LGI) by focusing on the simplest Leggett-Garg (LG) measure $K_3$.

We have introduced several improvements that are essential for a reliable analysis of real neutrino data.
First, we implemented a fully general treatment of the LG strings by considering all four possible sign assignments ($K_3^{++}$, $K_3^{+-}$, $K_3^{-+}$, and $K_3^{--}$) and defining $K_3^{\mathrm{max}}$ as the optimal measure, for both the quantum as well as the two classical (Factorized Correlator: FC, and Exponential Fit: EF) scenarios.
Second, we constructed energy triplets for each of the neutrino experiments directly from the widths of the energy bins themselves, which allowed us to include the effect of experimental energy bin width in the analysis.
Third, we introduced a split-Gaussian sampling scheme to propagate the asymmetric uncertainties present in the muon neutrino survival probability data.

We also showed that the fraction of LGI violations ($f_{\rm LGV} \equiv N_{\rm LGV}/N$) measure is generally not statistically well behaved for long-baseline experiments, as the classical distributions are typically sparse and non-Gaussian. To address this, we introduced a modified definition of the $z_{\rm RMS}$ measure, which is sensitive to both the number of energy triplets violating LGI as well as the magnitude of the said violation. Further, for experiments where the classical distributions exhibit non-Gaussian behaviors, we developed an effective CDF-based fitting method by applying a Gaussian CDF fit to the upper half of the empirical distribution.

The combination of $(i)$ considering the four sign assignments of the $K_3$ measure, $(ii)$ the modified $z_{\rm RMS}$ definition, $(iii)$ energy triplet determination from the experimental data, $(iv)$ split-Gaussian sampling of uncertainties, and $(v)$ an effective CDF fit procedure provides a general, robust, and reliable framework for determining the degree of quantumness in the neutrino sector.
In~\cref{fig:zscore_all_expts}, we presented the $z_{\rm RMS}$ histograms for MINOS, T2K, NOvA, and DUNE, comparing the quantum scenario against the 2 classical cases. We note that all four experiments exhibit a clear and statistically significant departure from classical hypotheses.
Taking the most conservative result between the quantum and classical FC as well as quantum and classical EF scenarios, the statistical significance of quantumness for MINOS, NOvA, and T2K reads
\begin{equation}
    N_{\rm MINOS}^{(\sigma)} = 3.55\sigma \, , \quad N_{\rm NOvA}^{(\sigma)} = 5.48\sigma \, ,\quad N_{\rm T2K}^{(\sigma)} = 13.86\sigma \, ,
\end{equation}
and, projection for DUNE (for 5\% systematic error) is
\begin{equation}
    N_{\rm DUNE}^{(\sigma)} = 7.98\sigma \; .
\end{equation}
Note that our T2K result appears to yield the strongest probe of quantumness at neutrino experiments considered to date.

The methods developed in this work apply directly to current datasets and offer a well-defined strategy for future neutrino experiments to probe quantumness. While in this work, we derived the results for the 4 long-baseline neutrino experiments, the pipeline presented is also directly applicable to non-accelerator-based neutrino experiments such as JUNO~\cite{JUNO:2015zny, JUNO:2025gmd, JUNO:2025fpc}, which has recently presented first results.

\acknowledgments
We thank Markus Ahlers, Daniel Cherdack, and Sebastian Shetters for useful correspondence. 
The work of VB is supported by the United States Department of Energy under Grant No. DESC0025477.

\bibliographystyle{JHEP}
\bibliography{biblio.bib}

\end{document}